\documentclass[review,authoryear]{elsarticle}

\usepackage{subfigure}
\usepackage{multirow}
\usepackage{rotating}

\journal{Safety Science}

\begin{document}

\begin{frontmatter}

\title{How crowd accidents are reported in the news media: Lexical and sentiment analysis}

\author[hongo,komaba]{Claudio Feliciani\corref{correspondingauthor}}
\cortext[correspondingauthor]{Corresponding author}
\ead{feliciani@g.ecc.u-tokyo.ac.jp}

\author[tue]{Alessandro Corbetta}
\author[unsw]{Milad Haghani}
\author[hongo,komaba]{Katsuhiro~Nishinari}
\address[hongo]{Department of Aeronautics and Astronautics, School of Engineering, The University of Tokyo, 7-3-1 Hongo, Bunkyo-ku, Tokyo 113-8656, Japan}
\address[komaba]{Research Center for Advanced Science and Technology, The University of Tokyo, 4-6-1 Komaba, Meguro-ku, Tokyo 153-8904, Japan}
\address[tue]{Department of Applied Physics and Science Education, Eindhoven University of Technology, P.O. Box 513 5600 MB Eindhoven, Netherlands}
\address[unsw]{School of Civil and Environmental Engineering, The University of New South Wales, UNSW Sydney, NSW 2052, Australia}

\begin{abstract}
The portrayal of crowd accidents by the media can influence public understanding and emotional response, shaping societal perceptions and potentially impacting safety measures and preparedness strategies. This paper critically examines the portrayal of crowd accidents in news coverage by analyzing the texts of 372 media reports of crowd accidents spanning 26 diverse news sources from 1900 to 2019. We investigate how media representations of crowd accidents vary across time and geographical origins. Our methodology combines lexical analysis to unveil prevailing terminologies and sentiment analysis to discern the emotional tenor of the reports. The findings reveal the prevalence of the term ``stampede'' over ``panic'' in media descriptions of crowd accidents. Notably, divergent patterns are observable when comparing Western versus South Asian media (notably India and Pakistan), unveiling a cross-cultural dimension. Moreover, the analysis detects a gradual transition from ``crowd stampede'' to ``crowd crush'' in media and Wikipedia narratives in recent years, suggesting evolving lexical sensitivities. Sentiment analysis uncovers a consistent association with fear-related language, indicative of media's propensity towards sensationalism. This fear-infused narrative has intensified over time. The study underscores the potential impact of language and sentiment in shaping public perspectives on crowd accidents, revealing a pressing need for responsible and balanced reporting that moves beyond sensationalism and promotes a nuanced understanding. This will be crucial for increasing public awareness and preparedness against such accidents.  
\end{abstract}


\end{frontmatter}

\section{Introduction}
When a large number of people gather, for example to attend events or to move by public transportation, accidents could happen in absence of proper crowd management. The accident which occurred in Itaewon (Seoul, South Korea) during Halloween 2022 \citep{WikiItaewon} is probably among the best known in the recent past, but tragedies keep occurring on a regular basis, such as in Yemen (April, 2023) or El Salvador (May, 2023), while they remain unequally advertised by the press \citep{WikiCrushList}. Although from a local perspective crowd accidents are relatively rare, on a global scale they occur, on average, once a month \citep{Feliciani2023}. It is estimated that in the period from 2010 to 2019 alone, over 4,500 people lost their lives in crowd accidents \citep{Feliciani2023}. Beside the casualties alone, it is important to consider the consequences that crowd accidents have on survivors and emergency respondents. Both may suffer long-term psychological damages often made worse by the belief that they could have done something to help prevent the tragedy from occurring \citep{Telegraph2023}. Further, events happening in countries with a wide English press coverage (e.g., the USA, UK, and India) tend to attract the attention of the international media leading subsequently to discussions around the likely causes and responsibilities. The mental image that lay people have of crowd accidents is therefore often shaped by the media, responsible to convey an accurate and factual description of the events without dramatizing them \citep{Luegering2023}. \par
In addition, the topic of crowd motion and behavior is studied by a growing number of researchers \citep{Haghani2021}. Both empirical \citep{Haghani2018} and theoretical \citep{Yang2020} studies are carried out to understand the fundamental principle of crowd motion \citep{Corbetta2023} to help in the management of mass events and design pedestrian infrastructure. The prevention of crowd accidents is a common motivation for researchers engaged in this field and some of them are also involved in the role of experts when it comes to investigating the causes of an accident. Typically, researchers agree on the physical mechanisms leading to these deadly conditions ultimately causing fatalities in crowd accidents. For instance, it is generally known that densities above 10 people/m$^2$ are at high risk of resulting in casualties and injuries (although this is not a necessary condition and fatal accidents can occur at lower densities) \citep{Murosaki2002,Helbing2007,Helbing2012,Feliciani2021}. Further, it has been shown that shock-waves or instabilities within the crowd will enable the creation of voids in which people involuntarily stumble leading to a pile up that suffocates people on the ground \citep{Helbing2007,Helbing2012,Still2014,Feliciani2021} (although the detailed mechanism is usually specific for each accident and a generalization is not possible). \par
In the field of collective psychology the understanding of phenomena related to crowd accidents has also evolved over the years \citep{Challenger2009A,Challenger2009B}. For instance, the group mind theory proposed by Le Bon at the end of the 19th century \citep{LeBon1895} and claiming that panic is contagious and spread through the crowd is now unanimously rejected and, over the last few decades, a widespread consensus has built up on what occurs (or better, what does not occur) in crowd accidents. In particular, it is known that panic occurs only on rare occasions and it does not spread through the crowd (some people may be in fear at the same time, but this is an individual feeling related to the fact that they are all in a very dangerous situation) \citep{Aguirre2011,Jacob2008,Sheppard2006}. Also, helping behavior has been often documented during crowd accidents and researchers agree this should be considered as a typical condition among crowd accidents \citep{Cocking2014,Drury2007,Drury2009}. \par
However, although mechanisms occurring in crowd accidents are now sufficiently understood, their presentation remain a problematic aspect. Most importantly, presentation of crowd accidents is particularly relevant when it comes to the press and popular media because news articles are read by a large number of people and they shape the understanding that lay people have on crowd accidents. Nonetheless, even in scientific literature terms such as ``panic'' or ``irrationality'' are often used improperly thus showing that the misrepresentation problem is not exclusive to media outlets or popular press \citep{Haghani2019}. \par
Still, despite the relative consensus that a more accurate representation of crowd accidents is important and needed, little research has been done to understand what the existing problems are, especially in regard to popular press. For instance, there is no clear image showing what are the problematic terms and whether their use is on the rise or not. The words ``stampede'' and ``panic'' are often claimed to be improperly used, but they are not necessarily the only examples. This lack of research limits our understanding of the problem, and, without a clear understanding of the current situation, it may be difficult to define strategies to tackle the problem in the future. \par
Nonetheless, despite a general lack of research on the representation of crowd accidents, a few studies considered this aspect in popular media from a systematic perspective. Among them we could include the study by Rogsch et al. \citep{Rogsch2010}. In their work, the authors analyzed 127 accidents presented in the ``classical literature'' (newspaper, TV, etc.) and considered whether terms related to panic had been used correctly in the description (according to the definition by the authors). The authors showed that ``panic'' and ``stampede'' were often used incorrectly and ``crush'' should have been used more often to provide an accurate description. Further, \citep{Haghani2019} analyzed the use of ``panic,'' ``irrationality,'' and ``herding'' within the scientific literature. The authors observed that the definition provided by researchers for each term changes depending on the field of study and, in general, there is no agreement on the meaning and the context in which those terms were used. The study offers a clear image on the use of controversial terms in the scientific community, but the same question remains open in regard to popular media. The question was partially addressed by L{\"u}gering et al. \citep{Luegering2023}, who, through a survey, studied the mental image that lay people have of crowd accidents. In addition, the authors considered whether this image could be changed by using a different word in presenting the facts (``mass panic,'' ``mass accident,'' and ``mass disaster'' were considered). The authors show that lay people typically associate selfish and rushed behavior with crowd accidents (similar to what the same people reportedly associate to animal behavior) and believe inappropriate behavior by the attendees is typically to blame. Further, they concluded that by just using a different word, the mental image would not change, hinting that similar words lead to the same (inaccurate) image. More recently, Sieben \& Seyfried \citep{Sieben2023} interviewed 136 witnesses of the 2010 Love Parade accident in Germany. In their work, they state that witnesses often indicated  helping behavior to describe what was seen during the accident and did not agree that mass panic occurred (although fear was indicated as an intense feeling). Finally, witnesses observed both a combination of falls and being moved by transversal waves, both phenomena already discussed and known to occur during deadly accidents. Similar conclusions were found by Cocking \& Drury who interviewed survivors of the Hillsborough 1989 accident \citep{Cocking2014}. \par
In short, the literature reviewed above tends to show that lay people have a distorted perception of crowd accidents, often involving panic and heterogeneous behavior, with the exception of witnesses who nonetheless represent an insignificant proportion of the population. Whether this is possibly caused by an inaccurate representation in the media is not completely clear, since efforts in analyzing the representation in popular media have been limited. \par
In this work, we improve and extend the analysis from the above studies on the representation of crowd accidents and focus on popular media. In doing so, we employ state-of-the-art text analytics tools which have become available only over the last few years: lexical and sentiment analysis. We would like to point out that both lexical and sentiment analysis are mainstream methods in different research fields, although the combination in which they are used here is rather novel. In addition, the analyzed dataset covering 372 press reports on crowd accidents represents another important contribution of this work. Its size, organization, and quality is what allowed us to use modern techniques to examine representation of crowd accidents in the media. \par
In our analysis we will study how reporting of crowd accidents has changed over time and whether the purpose of gathering (e.g., religious festival, sporting event, political rally, etc.) influences reporting style. Here, we aim to provide a structured analysis to support the discussion on misunderstanding of crowd accidents among lay people and comment on the role that news outlets play in that regard. Our work is intended to increase the awareness of word choice among researchers and academics working in the field of crowds, provide training material to practitioners and reporters, and help in the creation of strategies to improve understanding of collective behavior in crowd accidents. \par
The manuscript is organized as follows. Section 2 provides an overview on the controversial words taking a central role in this work. Section 3 presents the methodology and the material on which our analysis is based. Section 4 presents the main outcomes in which lexical analysis is discussed first, followed by sentiment analysis and special considerations for Wikipedia pages. Discussion and conclusions complete the manuscript.

\section{Controversial words}
Before discussing our study, it is worth considering some formal definitions of the two controversial words that play a central role in this work: panic and stampede. Here, we will discuss their use in the context of crowd accidents, but both words are also employed in other contexts without creating controversy. It is, therefore, important to introduce the formal definitions of both words and understand their etymological origin to take a more informed stance regarding their use in the context of crowd accidents. Dictionary definitions, etymology, and controversial aspects related to research on crowds are summarized in \tablename~\ref{tab:controversial_words}.

\begin{sidewaystable}
\centering
  \caption{Dictionary definitions, etymology, and controversial aspects in relation to the words ``panic'' and ``stampede.'' The Merriam-Webster Dictionary is used for American English and the Oxford English Dictionary for British English. Etymological definition is based on The Oxford dictionary of English etymology \citep{Onions1966}, with contributions from The Merriam-Webster Dictionary.}
	\label{tab:controversial_words}
\begin{tabular}{|p{2cm}|p{10cm}|p{10cm}|}
  \hline
																			& Panic						& Stampede			\\
  \hline
	The Merriam-Webster Dictionary			
	& a) a sudden overpowering fright;	b) a sudden unreasoning terror often accompanied by mass flight; c) a sudden widespread fright concerning financial affairs that results in a depression of values caused by extreme measures for protection of property (such as securities)
	& a) a wild headlong rush or flight of frightened animals; b) a mass movement of people at a common impulse; c) an extended festival combining a rodeo with exhibitions, contests, and social events \\
	\hline
	Oxford English Dictionary
	& Of the nature of, resulting from, or exhibiting a sudden wild, unreasoning, or excessive state of fear or alarm.
	&	A sudden rush and flight of a body of panic-stricken cattle. \\
	\hline
	Etymology	
	& Originating from the Greek \textit{p\={a}nik\'{o}s}, literally ``pertaining to Pan.'' Pan is the name of a Greek deity part man and part goat, whose appearance or unseen presence caused terror and to whom woodland noises were attributed. \textit{Panic} entered the English language first as an adjective suggesting the mental or emotional state that Pan was said to induce. The adjective first appeared in print at the beginning of the 17th century, and the noun followed about a century later.					
	& Stampede originates from the American Spanish \textit{estampida} in connection with \textit{estampido} (crack or bang noise) and \textit{estampar} (to stamp). This later form originates from the German \textit{stampfen} (originally \textit{stampfp\={o}n} in Old High German) and also meaning ``to stamp.'' The first known use of stampede was in 1828. \\
	\hline
	Controversial aspects
	& Sometimes mistakenly used as a synonym of fear or to imply a sense of urgency. A common critic is that panic is used for tragedies or disorderly collective behavior without verifying whether a panicking state of mind could be actually recognized in people.
	& Generally used as a synonym of crowd accident regardless on the mechanism seen in people and/or whether fleeing was physically possible or not. Critics sometimes argue that stampede should be only used for sparse crowds where fleeing is possible and crush or alternative words should be preferred for accidents involving very dense crowds. \\
	\hline
\end{tabular}
\end{sidewaystable}

Regarding panic, it is worth noting that the Merriam-Webster definition already accounts for the possibility of mass flight. From this perspective, we can argue that a flight caused by a terror attack could be correctly described as related to panic. However, it is also important to note that no definition implies a mechanism of panic contagion, which is sometimes described or implicitly inferred when discussing crowd accidents. From this perspective, panic should be used in crowd accidents to describe a widespread but individually constrained ``excessive state of fear or alarm.'' However, the threshold for determining how frightening a situation should be to assume people are in a state of panic and not simply worried is open to discussion. \par
When considering stampede, it is worth noting that its definition is fairly similar to that of panic. The Oxford Dictionary indeed includes panic as a condition to define a stampede, although no reference to human crowds is made. The Merriam-Webster distinguishes between ``frightened animals'' and people moved by a ``common impulse,'' possibly implying less urgency in the context of human crowds. Regardless of these minor differences, both dictionaries agree in associating a stampede with a sudden rush resulting in a mass movement. From this perspective, only crowd accidents involving people running or fleeing should be described as a stampede, whereas accidents occurring in dense crowds without a rush would not fall into the definition of a stampede. \par
In conclusion, it is interesting to note that both panic and stampede are associated with animal behavior. Panic originates from the Greek god Pan: the god of the wild, shepherds, flocks, rustic music, and impromptus \citep{Brown1981}. The term ``stampede'' is also used to describe events, including those in a rodeo. The question of whether animal behavior should be used to describe humans (and vice versa) and to what extent we can compare human and animal behavior is an open debate \citep{DeWaal2016}. It is therefore possible that part of the criticism regarding the use of panic and stampede for human crowds stems from the relation that both words have to the animal kingdom.

\section{Data collection and analytical methods}
In this section, we can now turn into more technical aspects of this work. Here, we will present the characteristics of the dataset on which this work is based and discuss the methods employed in its analysis. Since analytical tools are derived from existing works, we will mainly focus on describing the dataset and briefly describe the methods while providing relevant references to interested readers.

\subsection{Data collection}
\subsubsection{Press reports}
The corpus made of press reports collected in the frame of this study certainly represents one of the most important elements upon which results are generated and discussed. However, presenting the details of the method employed to select reports and its sources requires an extensive discussion. Regardless, important knowledge needed to understand the results can be outlined through a short presentation. We will therefore provide relevant details here and address readers looking to delve into minor methodological aspects to Appendix A. \par
In short, there are two challenges encountered when attempting to create a representative corpus of press reports. Reports need to be large in number, come from a variety of sources, and yet allow to represent nuances in reporting style over time and geographical area. We therefore decided to focus on media outlets having a long tradition and which covered a large number of accidents. As a result, the BBC, New York Times, The Guardian, The Times of India, etc. were selected. In total 372 press reports were obtained from 26 sources covering 188 different crowd accidents. \par
For each report we extracted the reporting institution, the source (typically in the form of a URL), accident and publication date, the title (or the headline), and the content of the report (i.e., the body). These information have been used to categorize each report and provide the text used in both lexical and sentiment analysis. Due to copyright restrictions in relation to most of the sources, texts cannot be shared. Nonetheless, URL of the sources and aggregated material from the analysis is available at \citep{ZenodoDataset2023B} to researchers interested in alternative analyses.

\subsubsection{Wikipedia pages}
Although traditional mainstream press is the main focus of this work, Wikipedia pages also represent a relevant and interesting source of information. In fact, Wikipedia is commonly used to learn about previous historical events and, in contrast to media reports, it can be modified over time through collaborative work. Its format and structure consequently allows it to track changes over time and detect minimal changes in word use and associated frequencies. On this scope we automatically scraped the content of Wikipedia pages describing the accidents contained in our list. More precisely the pages relative to 68 accidents were scraped through a dedicated script on two occasions: on October 15th, 2022 and again on May 25th, 2023 (reasons for this choice will be explained when presenting the results). Similarly to media reports only title and body were extracted, thus excluding subtitles or captions.

\subsubsection{Data on English literature and scientific publications}
As already discussed in the introduction, the topic of word use in relation to crowd (accidents) within the scientific literature has been already covered and an exhaustive discussion is for example provided in \citep{Haghani2019}. Yet, a comparison between use in popular media and alternative sources could help enrich the discussion. For these reasons, we decided to collect aggregated information on the use of controversial words in the scientific and general English literature (books, regardless of the subject area). Data for books were obtained from the Google Ngram service \citep{Michel2011}, while scientific literature was analyzed based on the Web of Science database.

\subsection{Analytical methods}
\subsubsection{Grouping approach}
The approach presented above allowed us to collect a sufficient number of texts to be analyzed through statistical and machine learning methods that will be presented in detail below. However, our interest is in recognizing trends in reporting of crowd accidents and therefore the whole dataset needs to be partitioned into different subsets of reports to be cross-compared. In this work we specifically focus on three different attributes associated to each report: 1) publication year, 2) geographical origin and 3) purpose of gathering. Each attribute is further explained as follows.

\begin{itemize}
	\item \textbf{Publication year:} To recognize temporal trends in reporting styles, reports were divided into five time periods: 1900-1979, 1980-1999, 2000-2009, 2010-2014, and 2015-2019. The uneven distribution of each period can be explained by the increasing number of crowd accidents reported over the last 20 years. The chosen distribution allows us to strike a balance between a sufficient sample size in each time period and the recognition of temporal trends.
	\item \textbf{Geographical area:} To study whether there is a tendency to use some words in specific geographical areas we grouped each source into four macroscopic regions: Europe (including UK); India and Pakistan; USA; and other (which includes all other countries, e.g. Australia). Please note that the geographical area refers to the location of the reporting institution and not to the location of the accident. Accordingly, a report by The New York Times on an accident in India will be classified as originating from the USA.
	\item \textbf{Purpose of gathering:} Each report was categorized taking into account the type of gathering related to each accident. The following categories were used in line to the analysis presented in \citep{Feliciani2023}: application (e.g., applying for a public position), donation, educational (accidents which occurred in schools), entertainment (e.g., a concert), political (e.g., a rally), religious, shopping (e.g., Black Friday sale), sport (soccer in particular), transportation (accident occurring in train stations or similar facilities), and other/unknown.
\end{itemize}

Using the above definitions, reports were combined into text corpus which were compared within a specific context using the methods explained below.

\subsubsection{Text analysis}
The analysis of large volumes of text is becoming an increasingly popular research subject given the amount of information openly shared through the internet. Scientific publishing is particularly suited to this type of research because databases containing scientific literature are well maintained and contain a number of auxiliary information (e.g., references) allowing the creation of so-called bibliometric networks. Bibliometric networks rely on references from scholarly publications to link authors and works forming a network of connections that can be analyzed to understand how science evolves through collaboration and citations \citep{Yan2012,Batagelj2013}. Bibliometric networks allow, for example, to determine how related different research fields are (e.g., medicine is more closely related to psychology compared to geology) or to understand how researchers collaborate with fellow colleagues. \par
Although the study of bibliometric networks is becoming an increasingly specialized area of research, the techniques developed by researchers in that field are not specific to it and lie on theories derived from network science and statistics. We can therefore apply the same techniques to analyze the media reports contained in our dataset. In this work we used the VOSviewer software \citep{VanEck2010} which is gaining popularity for scientific bibliographic analysis, but also offers the possibility to analyze the corpus of texts such as the dataset presented above. In the following discussion, we will present how the algorithms implemented in VOSviewer have been implemented in our work. However, here, we use only a subset of them, so readers interested in more details are addressed to the project website \citep{VOSviewer} or its original publication \citep{VanEck2010}.

\begin{figure}[htbp]
		\centering
		\subfigure[{Binary count} 	\label{fig:example_binary}]
		{\includegraphics[width=92mm]{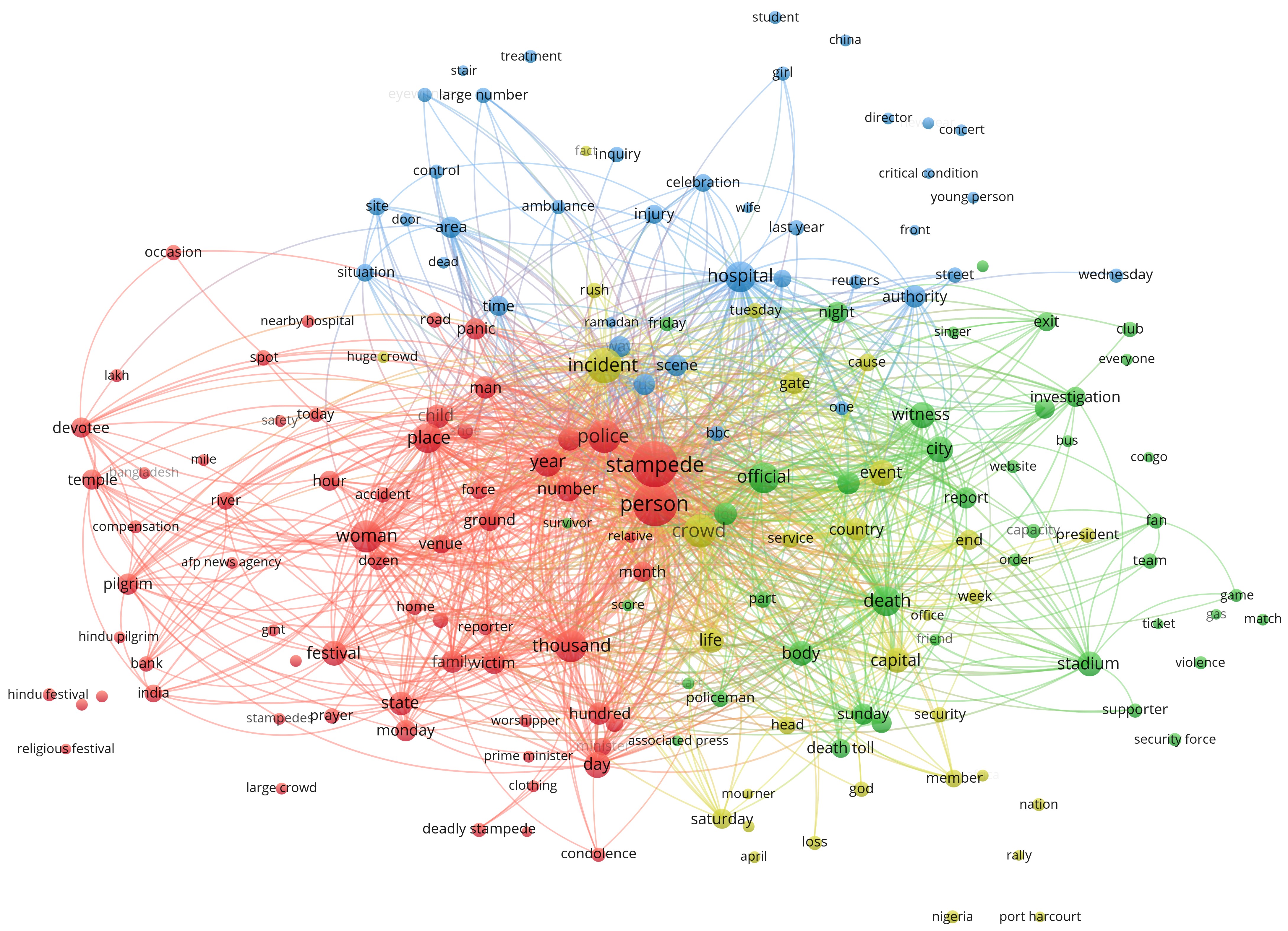}}
		\subfigure[{Full count} 			\label{fig:example_full}]
		{\includegraphics[width=92mm]{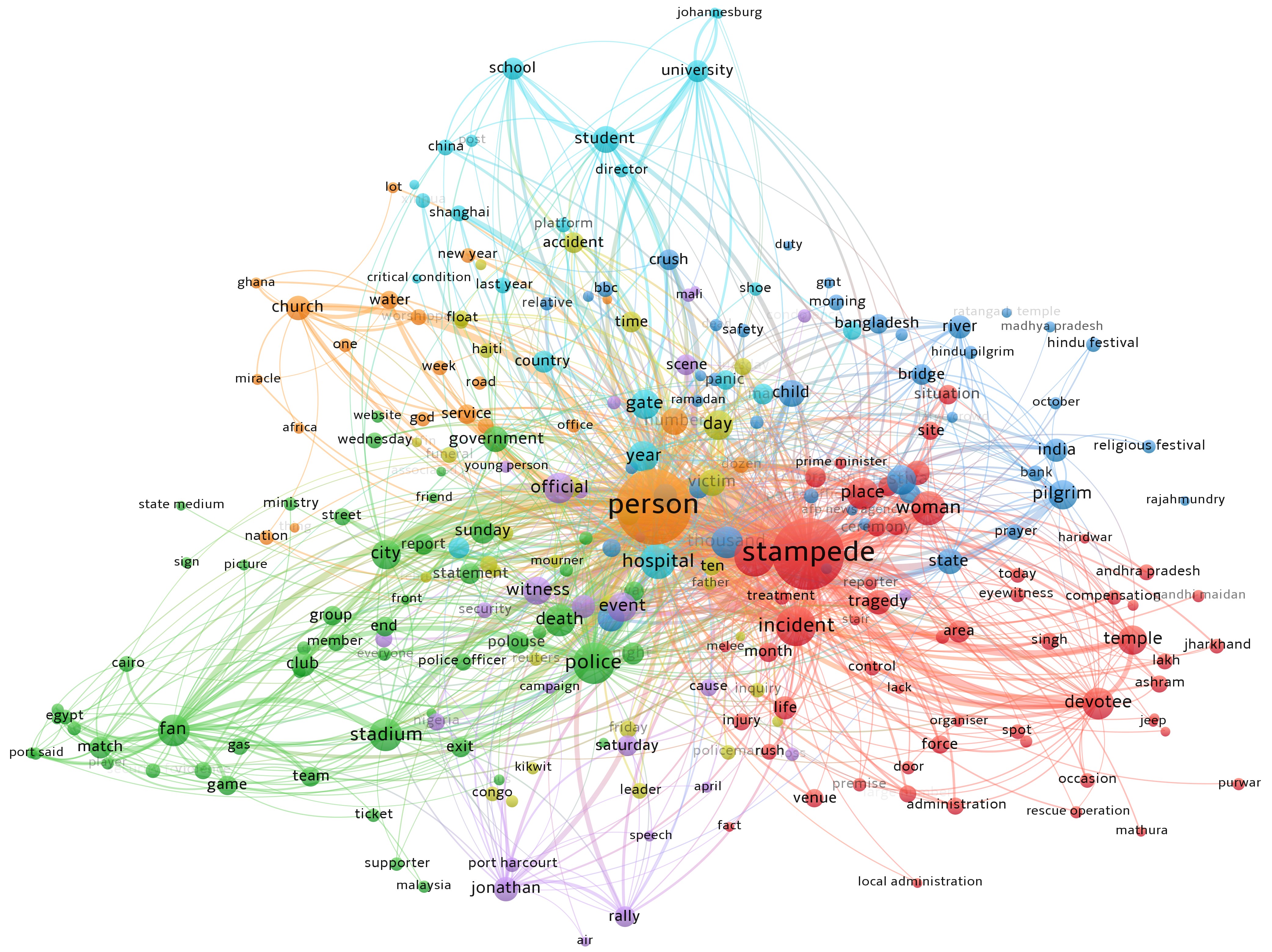}}
    \caption{Text maps generated using VOSviewer through different approaches. Both maps are relative to reports published in the period between 2010 and 2014. This specific subset was chosen since it clearly highlights differences between the binary and full count approach. Only words that have been used more than five times \textit{in total} are used in the analysis to simplify the visualization and exclude words that play a negligible role. Size of each word is associated with its occurrence (in the whole subset of reports from 2010 to 2014). Link strength is associated with co-occurrence. Color and positions are assigned through a clustering algorithm.}
		\label{fig:vos_example}
\end{figure}

To focus on the elements which are relevant to this work, we may consider a result generated through VOSviewer and use this example to discuss the properties of each text map and what can be extracted from it. \figurename~\ref{fig:vos_example} presents two text maps relative to all accidents considered in our dataset reported between 2010 and 2014 analyzed using two different statistical approaches. Before getting into statistical aspects it should be noted that VOSviewer automatically considers the syntactical structure of the English language and thus articles or prepositions are excluded from the analysis and plurals are converted into singular words to allow a rigorous analysis. This can explain why generic words like ``the'' are not present and why ``person'' (the singular of ``people'') appears centrally in both plots of \figurename~\ref{fig:vos_example}. We should further add that the title and body were combined to create the various corpus studied through lexical analysis.

\paragraph{Binary vs. full count} \par
In general, what is relevant for our analysis is to know how often a word has been used in a specific context, e.g., a time period in this example, and which pair of words are more commonly used within the same report. To perform this kind of analysis there are two different strategies which are possible. In the binary approach it is only considered whether a word is used or not in a specific report. Accordingly, 0 is used when a word is not found and 1 whether it is found regardless of the number of occurrences. In contrast, in the ``full count'' approach the number of times a word is used is accounted for, thus resulting in a discrete range of values ranging from 0 up to the maximum number of counts. \par
Based on this fundamental information, VOSviewer later computes occurrences of each word within a specific context (again, the time period between 2010 and 2014 in this example) and the co-occurrences between each combination of words. Occurrences simply represent the number of times a word has been used in total within a specific corpus of text. In the word map, occurrences are visualized through the size of each word, with words being used more commonly appearing larger. Co-occurrence between two words shows how commonly a set of words has been used within the same report. Co-occurrences are visualized through the weight of each link connecting a set of words. Further, specific algorithms are used to position words in the bi-dimensional space and assign a color through a clustering method. Although position and color are helpful to gain a qualitative understanding, they represent a high-dimensional projection of occurrence and co-occurrence and they are partially dependent on the clustering algorithm employed. For these reasons, in our analysis, we will simply make use of occurrences and co-occurrences which are easy to understand and can be directly associated with trends in the use of specific words. \par
In \figurename~\ref{fig:vos_example}, it can be seen that binary count and full count partially lead to quantitatively different but qualitatively similar results. For instance, words like ``stampede,'' ``person,'' ``police,'' or ``incident'' are associated with high occurrences in both representations. Due to the unbounded approach taken in the full count approach, differences between words tend to be larger and this helps the clustering algorithm in producing an appealing image where it is possible to recognize words associated with religious events (red) or soccer matches (green). However, the full count approach is partially biased by the length of each report. Longer reports lead to an increased use of all words, regardless of whether a word is used often because it plays a central role or simply because a report is particularly long. From this perspective the binary approach is closer to the scope of this work, but its statistical approach (i.e., considering occurrences only binarily) could have an impact on the results, especially when co-occurrences are computed. In other words, both the binary and full count approach have advantages and disadvantages but yet both represent a valid approach to analyze the use of words in press reports. For this reason we decided to employ both approaches and results will be consequently presented using the binary and full count approach.

\paragraph{Occurrence vs. total link strength} \par
Considering the different subset of texts relative to each category a large number of word maps can be generated making a structured comparison difficult. To simplify the comparisons, we will therefore only focus on occurrences and co-occurrences relative to each word map. Occurrences can be directly obtained from VOSviewer for each word, but co-occurrences are relative to combination of words. We therefore compute what we will define as the ``total link strength,'' i.e., the sum of all co-occurrences associated with a specific word. Visually this corresponds to summing up the weight of all links connecting a word with others in \figurename~\ref{fig:vos_example}. In more practical terms, we can associate occurrence with the popularity of a specific word and total link strength with the importance played within the corpus of text. As we will see both often lead to similar results. \par
Despite the simplification performed in this work, the original word map for each approach (binary or full count) and for all categories can be nonetheless interactively consulted online at \citep{ZenodoDataset2023B}.

\subsubsection{Sentiment analysis}
Although the frequency in which words are used can help recognize trends, we should also not forget that words rarely carry a neutral message. In general, every word can be associated with a particular emotion, which can be usually classified as positive or negative. With the recent improvements in natural language processing and machine learning it is now possible to estimate with a sufficient degree of accuracy emotions contained in specific messages. For example, ``I feel sad for having said that'' may be associated with a feeling of sadness which is generally categorized as a negative emotion. ``Thank you so much!'' can be associated with joy, being a positive emotion. Although we expect that the description of crowd accidents will be overwhelmingly associated with negative feelings, the circumstances of an accident may trigger very specific emotions. For instance, emotions associated with tragedies occurring at religious festivals may be different if a similar accident occurs during a soccer game. \par
In this work, we employ the DistilBERT model \citep{Sanh2019,DistilBERT} to detect the different spectra of emotions associated with different types of accidents \footnote{DistilBERT is a ``general-purpose language representation model'' \citep{Sanh2019} created to handle tasks associated with Natural Language Processing (NLP). In this work, we utilize a specialized version of the DistilBERT model specifically designed for extracting emotions from short texts \citep{Savani2023}.}. For a specific text the model can provide levels related with different types of sentiments generally associated with the wheel of emotions (the concept was originally proposed by R. Plutchik \citep{Plutchik1982} although a large number of modifications were proposed over the years, e.g., the ``hourglass of emotions'' \citep{Cambria2012}). Specifically, the model employed here allows us to estimate love, joy, surprise, sadness, anger, and fear. \par
Each text was evaluated through the model and emotions were later averaged for each category considered in our analysis. Since the model has a restriction of 250 characters per text, longer texts were split into shorter sentences and the overall result was taken by averaging all sentences. For each portion of text, emotions were extracted using the algorithm, resulting in a distribution over the six possible outcomes (e.g., love, anger, fear, etc.). However, in line with good practices in machine learning, only the emotion with the highest score was retained and assigned a value of 1. Averages were calculated for long texts by combining the results of short sentences, and later, a full average over the corpus of text was taken. Finally, in sentiment analysis, we excluded reports published more than one year after the accident because emotions may differ from reports published as breaking news.

\section{Results}
This section presents the main outcomes of this study. We will start by discussing lexical analysis and then move on to sentiment analysis. For both approaches we will consider, in order, historical trends, geographical origin of the source, and purpose of gathering. The section will be completed by a special discussion on Wikipedia reports before and after the Itaewon accident.

\subsection{Lexical analysis}
Lexical analysis will be presented using simplified diagrams making interpretation of the results easier, but neglecting some information (such as the numerical values for occurrences and total link strength associated with each word). For completeness, tables including those data are presented in Appendix B and raw data are shared in \citep{ZenodoDataset2023B}. It is worth remembering that, unlike the sentiment analysis presented next, in the lexical analysis, title and body was combined in computing the results.

\subsubsection{Historical trend}
\begin{figure}[htbp]
		\centering
		\subfigure[{Binary count occurrence} 						\label{fig:binary_time_occurrence}]
		{\includegraphics[height=46mm]{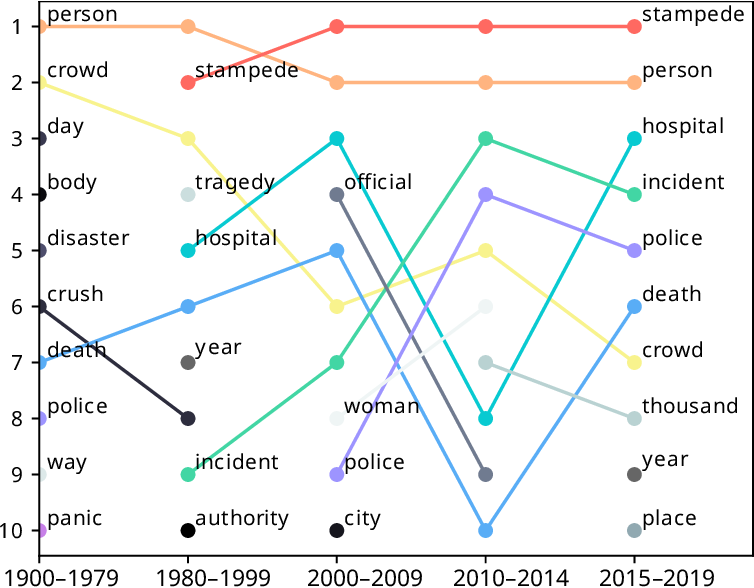}}
		\subfigure[{Binary count total link strength} 	\label{fig:binary_time_strength}]
		{\includegraphics[height=46mm]{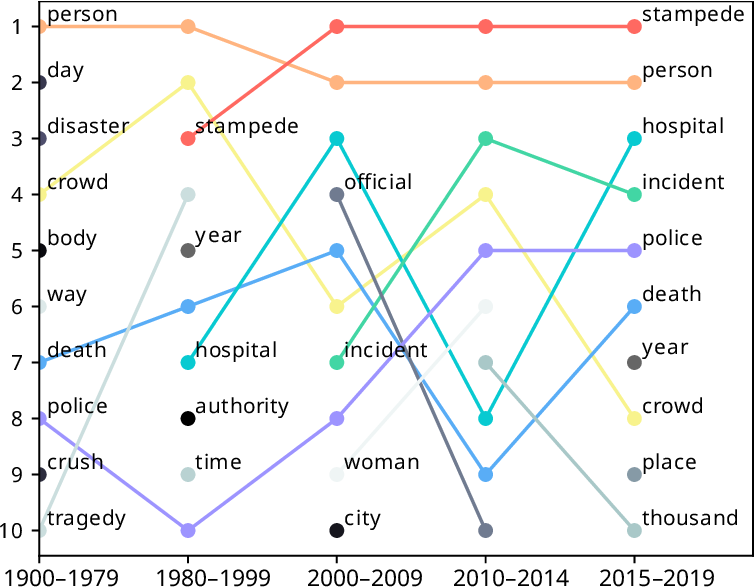}}
		\subfigure[{Full count occurrence} 							\label{fig:full_time_occurrence}]
		{\includegraphics[height=46mm]{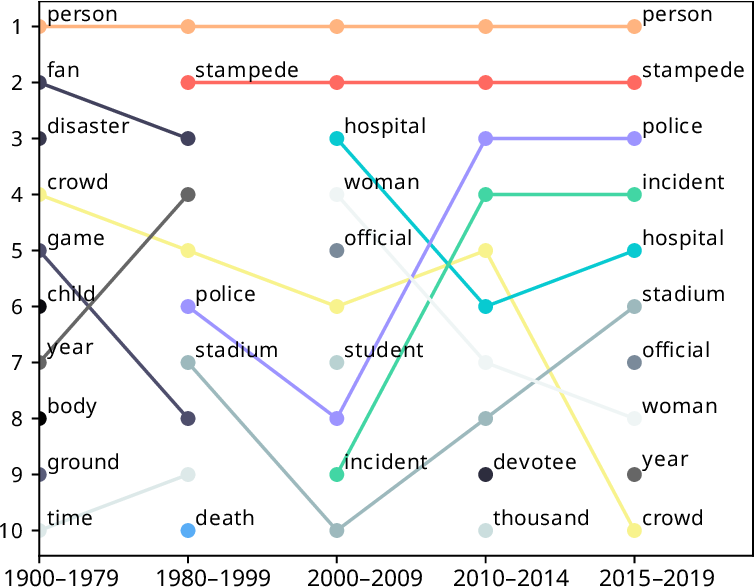}}
		\subfigure[{Full count total link strength} 		\label{fig:full_time_strength}]
		{\includegraphics[height=46mm]{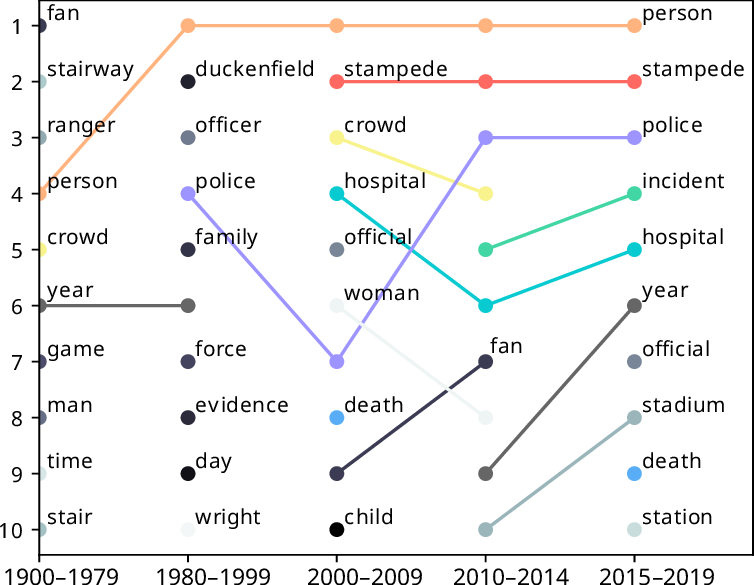}}
    \caption{Rankings showing how the use of words in crowd accidents reports have changed over the last 120 years. For each period only the top ten words are visualized and different interval sizes are used to ensure enough samples in each period.}
		\label{fig:words_time}
\end{figure}

We begin by considering historical trends in the use of words through the lexical analysis of our dataset. \figurename~\ref{fig:words_time} shows the changes through the considered time periods in respect to occurrence and total link strength. To simplify the interpretation of word maps, we computed the word ranking for each measure in each time period using both the binary and full count method. Words that are used the most or play a central role in the text will result in a higher position in the ranking. Lines are used to link the same words over time periods.

Regardless of the method (binary vs. full) and the measure (occurrence vs. total link strength) used we can observe that the top two most commonly used words are ``person'' (more likely used in the plural form of ``people'') and ``stampede.'' In the binary count it can be seen that ``person'' and ``stampede'' exchanged their position around the year 2000. In the full count, although positions remain constant, we can observe that ``stampede'' starts appearing more or less in the same period, i.e., around 2000. Except for ``person'' and ``stampede'' there are only a few other words that constantly appear in the ranking. For instance, ``crowd'' appears to play a less important role, taking top positions in the first time period and appearing at the bottom during the more recent one. On the other hand, the word ``incident'' appears to be more increasingly used, especially after 2010. ``Hospital'' also appears commonly among the top ten words although there is no clear trend that can be observed. The word ``death'' appears in all periods in the binary count, but the same cannot be seen in the full count. This means that ``death'' is often used at least once, but may get overshadowed by other words possibly used in repetition when the full count is employed. Finally, the word ``police'' also appears to be increasingly used, and, despite a trend that is difficult to recognize, it is among the top-five most common words for the period 2015--2019 (regardless of the approach).

\paragraph{Qualitative comparison with alternative sources} \par
To contextualize the results presented above we can also compare the use of specific words in different corpus to come up with some hypothesis explaining their usage and diffusion. Discussing the use of the word ``person'' would be rather pointless since it is used in any context and is not problematic regarding the representation of crowd accidents. However, ``stampede'' has been often portrayed as a problematic word and its use is very specific to crowd accidents (with the partial exception of animal behavior where it is also used). In this analysis we will also consider the word ``panic,'' which, although being the arguably most controversial term in regard to crowd dynamics, was almost irrelevant in the presented rankings (it only appears in \figurename~\ref{fig:binary_time_occurrence} at position 10 for the period 1900--1979).

\begin{figure}[htbp]
		\centering
    \includegraphics[width=60mm]{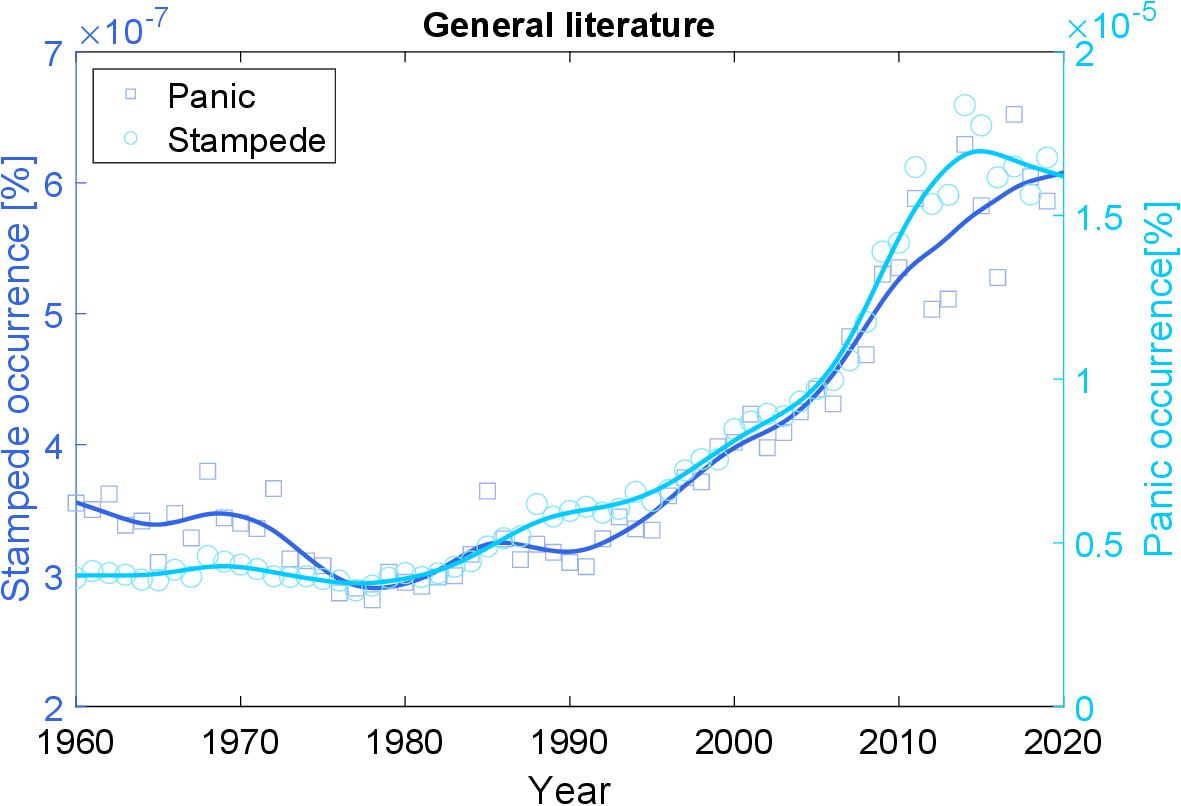}
    \caption{Use of the words ``stampede'' and ``panic'' in the general literature represented by the Google books corpus \citep{Michel2011}. A smoothing spline is used to show the trend among the dots for each word. Starting year is taken as 1960 to show the changes in trend over the last six decades. Relative use (i.e., percentage) of each word relative to the whole corpus is used with a different scale for both words.}
		\label{fig:trend_ngram}
\end{figure}

Our discussion starts by considering the general literature as represented by the corpus of Google books. Usage of ``stampede'' and ``panic'' are represented in \figurename~\ref{fig:trend_ngram} for the period from 1960 to 2020. Despite showing a somewhat similar trend (with minor differences) we should consider that ``panic'' is generally around 20 times more common. This can be explained considering the wide use of ``panic'' in different fields, especially psychology (e.g., panic disorder). Yet, we can observe that the relative use of the word ``panic'' has grown from 1980 but generally slowed down over the last 5 years (the peak appears in 2014). On the other hand, ``stampede'' has regained popularity from the 1980s and, although the trend is not crystal clear, its use seems to be on the rise.

\begin{figure}[htbp]
		\centering
		\subfigure[{Scientific publications (overall)} 		\label{fig:scientific_publications}]
		{\includegraphics[width=60mm]{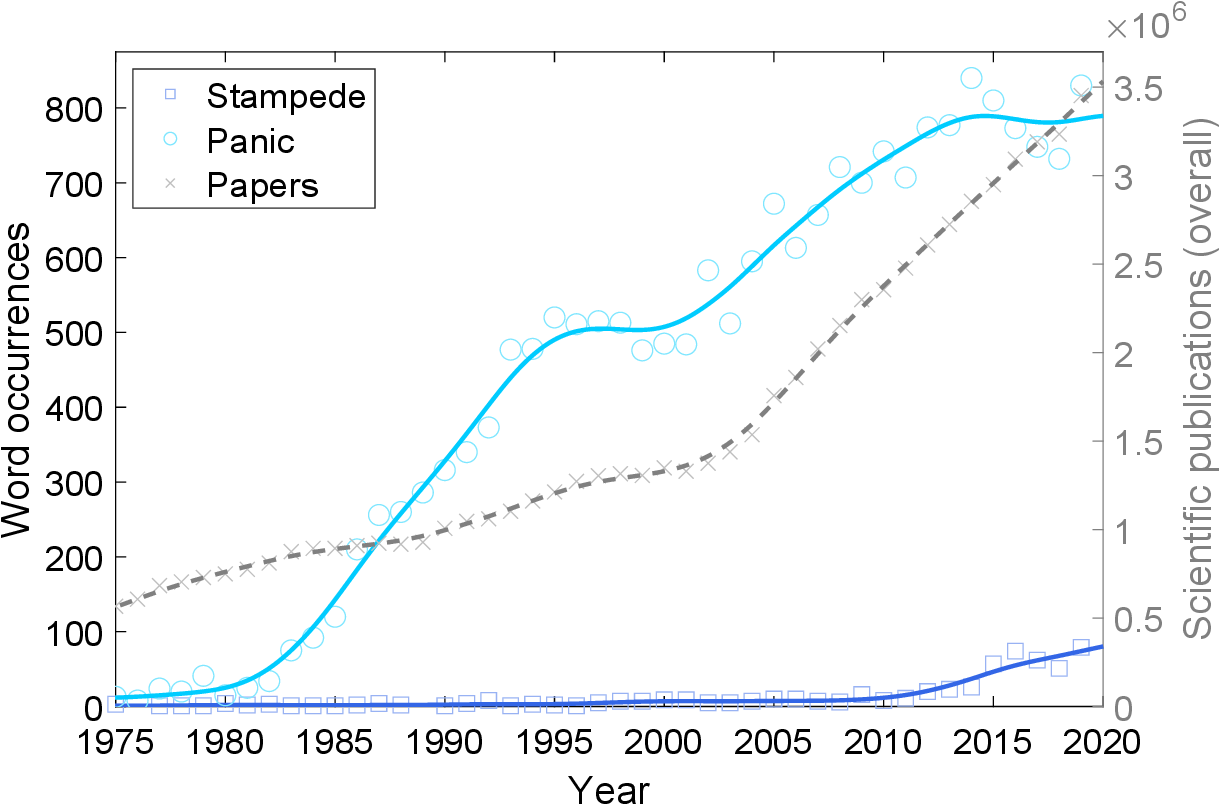}}
		\subfigure[{Publications on crowd dynamics} 			\label{fig:scientific_crowd}]
		{\includegraphics[width=60mm]{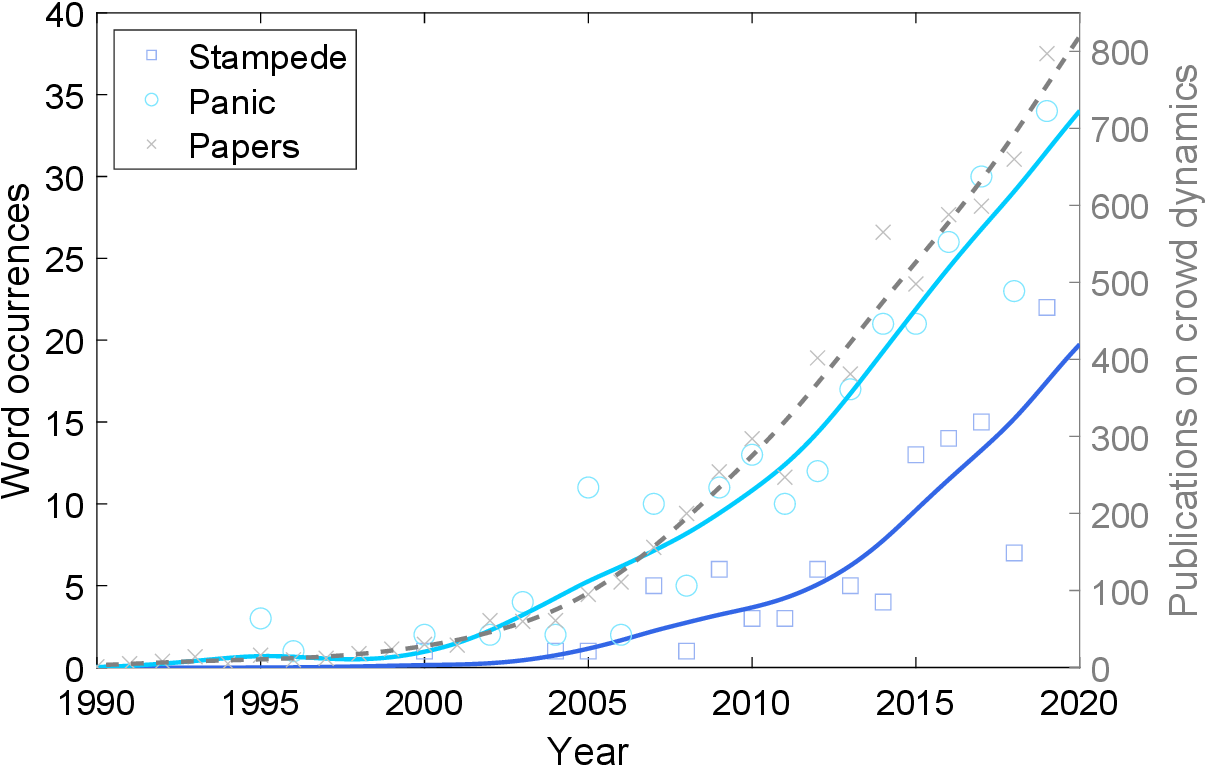}}
    \caption{Use of the words ``stampede'' and ``panic'' in scientific publications listed in the Web of Science (WoS) database. Starting year is set to provide an overall view on the evolution and yet consider limitations of the databases (coverage of the WoS is only partial for older publications and crowd dynamics was barely studied before 2000).}
		\label{fig:papers_trend}
\end{figure}

Although the general literature allows us to see when trends emerged, especially for ``stampede;'' we also need to acknowledge that the range of texts covered is very different: going from novels to technical manuals or biographies. Thus, we will consider the corpus of scientific publications in the Web of Science (WoS), which, despite being diverse among disciplines, only contains non-fiction texts. The evolution of ``stampede'' and ``panic'' in the WoS is presented in \figurename~\ref{fig:scientific_publications}. In the WoS too, ``panic'' is used more widely than ``stampede,'' although we can possibly attribute this finding with its use in the frame of psychology/medicine. Panic disorder was first introduced in the DSM (Diagnostic and Statistical Manual of Mental Disorders) in its third edition in 1980 \citep{DSM1980,Nardi2020}. This can explain the consequent rapid rise in the use of ``panic.'' On the other hand, the use of ``stampede'' appears to rise from 2010 onward (it appeared in less than 10 publications per year before 2005). This is remarkably later than the first appearance in the top-ten of press reports in the period around 1980--1999 (see \figurename~\ref{fig:words_time}) and the changing trend in the general literature observed around 1990 (see \figurename~\ref{fig:trend_ngram}). We can therefore speculate that the use of ``stampede'' in scientific publications followed an increase observed in the general literature and possibly, partially, by its use in reporting crowd accidents. \par
In order to better understand the evolution of the word ``panic'' we need to restrict the scope of our investigation. \figurename~\ref{fig:scientific_crowd} presents the use of both ``panic'' and ``stampede'' within scientific publications covering the topic of ``crowd dynamics'' (its working definition is provided in \citep{Haghani2021}). As can be observed, the usage of both words can be associated with an increasing number of publications on crowds. However, here, we need to remind readers that panic has been rarely seen in crowds and therefore its use in crowd-related publications can be rather puzzling. As described in details in \citep{Haghani2019}, the increasing use of ``panic'' can be related to an influential publication from 2000 \citep{Helbing2000}. Although this work introduced important elements helping in the simulation of crowd motion, it also led to a misleading use of ``panic'' within the scientific community working on crowds. \par
As a side note, it should be mentioned that despite the multiple perspectives considered in the above analyses, the intention was not to infer the correctness of the terms from this trend. Whether each word has been used correctly or not can be concluded only by examining each text. However, considering that stampede and panic \footnote{Here stampede and panic are intended as behavioral features, not words used in a text.} are generally rare in fatal crowd accidents (especially the latter) we can assume that these words should rarely appear in the representation of such accidents. Under this assumption we can therefore argue that ``stampede'' is a generalized problem with the word being used in a range of contexts: from scientific literature on crowds to press reports covering accidents. On the other hand, the misrepresentation in relation to the word ``panic'' appears to have been partially amplified by an incorrect and frequent use within the scientific literature on crowds.

\subsubsection{Geographical origin}
\begin{figure}[htbp]
		\centering
		\subfigure[{Binary count occurrence} 						\label{fig:binary_area_occurrence}]
		{\includegraphics[height=48mm]{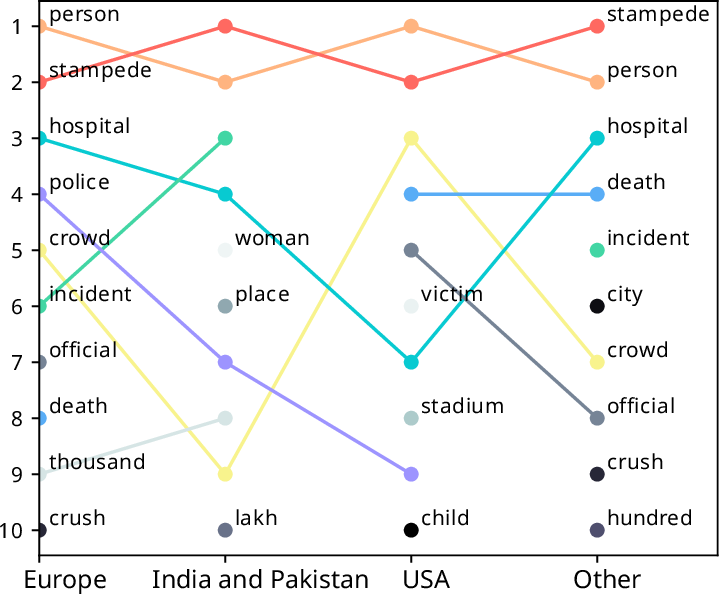}}
		\subfigure[{Binary count total link strenght} 	\label{fig:binary_area_strength}]
		{\includegraphics[height=48mm]{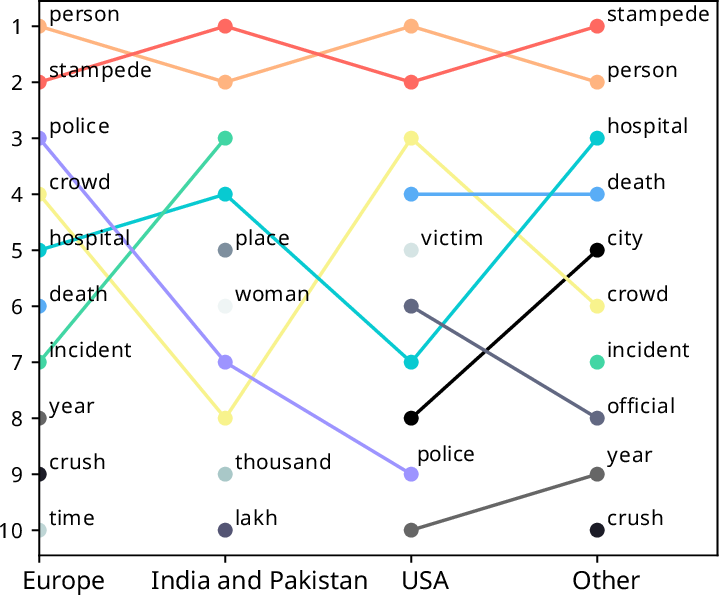}}
		\subfigure[{Full count occurrence} 							\label{fig:full_area_occurrence}]
		{\includegraphics[height=48mm]{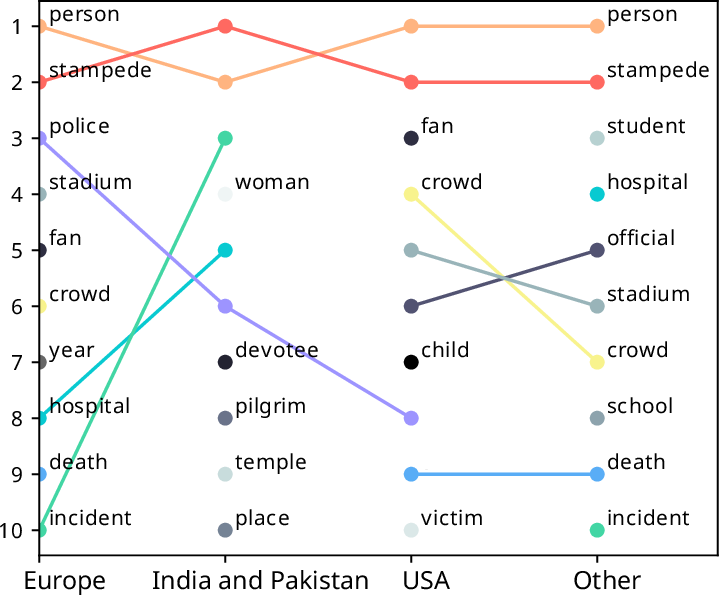}}
		\subfigure[{Full count total link strength} 		\label{fig:full_area_strength}]
		{\includegraphics[height=48mm]{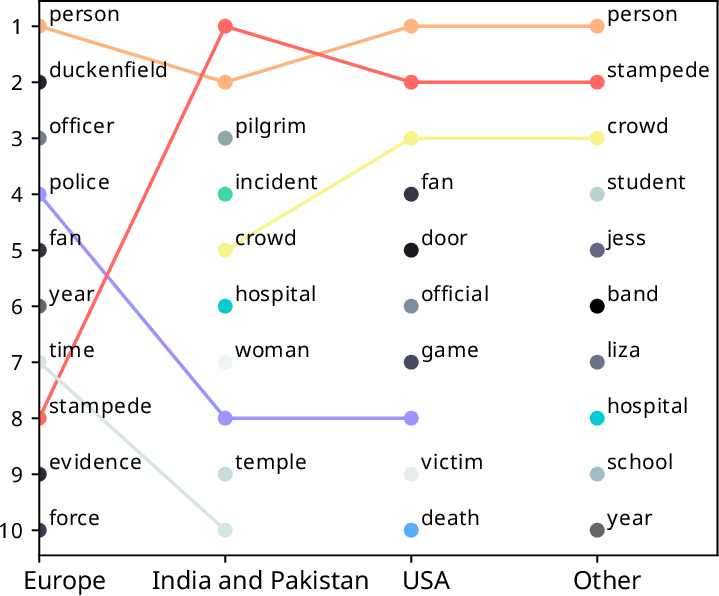}}
    \caption{Use of words in crowd accidents media depending on the geographical location of the reporting institution. Lines are used to link and highlight words that are frequently used although no trend is being visualized here.}
		\label{fig:words_area}
\end{figure}

We can now proceed with the analysis of media reports by considering their geographical origin. Results are presented in \figurename~\ref{fig:words_area}.

Similar to the historical trend, ``person'' and ``stampede'' are constantly appearing among the most used words. However, the use of ``stampede'' appears to be slightly less common in Europe and the USA, although it still remains a very common word. Further, an interesting difference is found for the words ``crowd'' (or ``police'') and ``incident.'' Both ``crowd'' and ``police'' take top position in European media reports and are also among the top-ten words for the USA. But in India and Pakistan and, even more, in the rest of the world they are less frequently employed. On the other hand, the word ``incident'' appears in a higher position for Indian and Pakistan media. As we will see through the sentiment analysis, there is a tendency from Indian and Pakistan media to report the news more dramatically. \par
Finally, we can see that, especially in the full count, words used in each region are reflective of the gatherings occurring there. In Europe the words ``fan'' and ``stadium'' appears in the top-ten, whereas ``devotee,'' ``pilgrim,'' or ``temple'' only appears in India and Pakistan. Words like ``school'' and ``student'' can be associated with a number of accidents in schools in China (included in ``other'' in our categorization).

\subsubsection{Purpose of gathering}
The lexical analysis will be completed by considering the use of words in relation to the purpose of gathering where accidents occurred.

\begin{figure}[htbp]
		\centering
		\subfigure[{Binary count occurrence} 						\label{fig:binary_event_occurrence}]
		{\includegraphics[height=48mm]{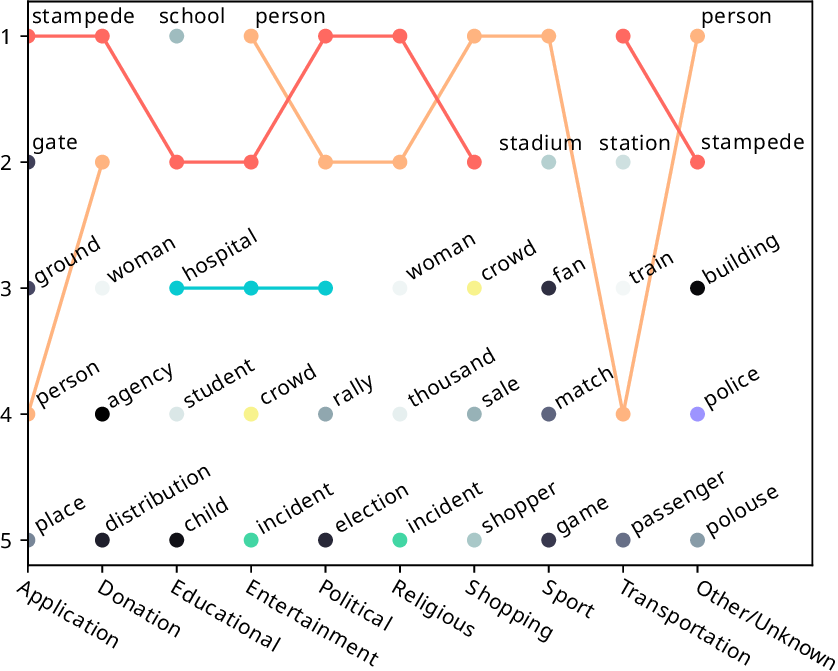}}
		\subfigure[{Binary count total link strength} 	\label{fig:binary_event_strength}]
		{\includegraphics[height=48mm]{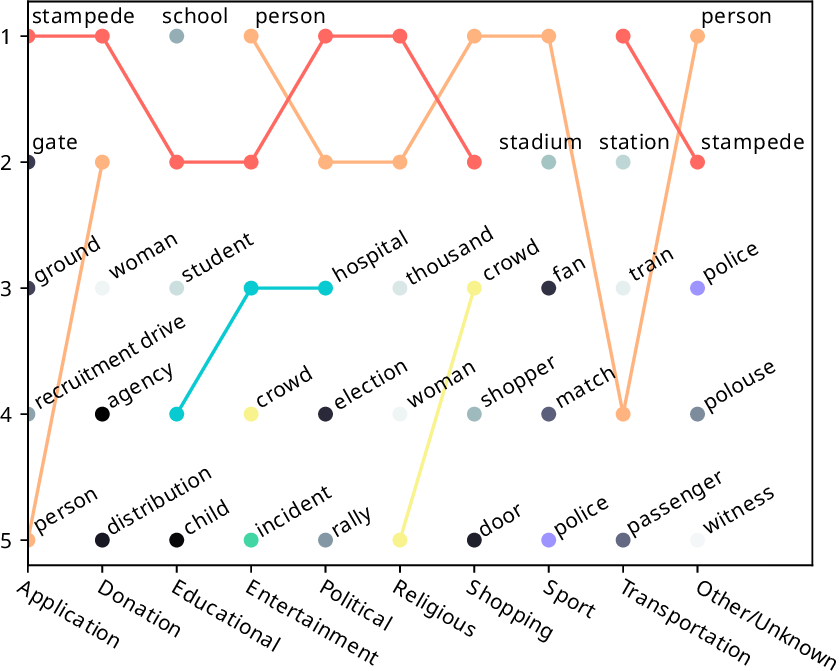}}
		\subfigure[{Full count occurrence} 							\label{fig:full_event_occurrence}]
		{\includegraphics[height=48mm]{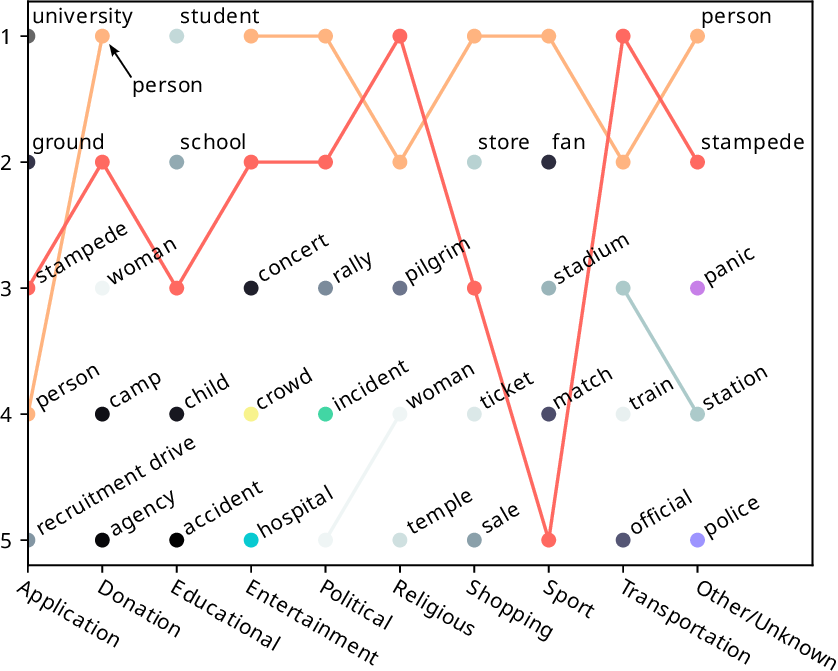}}
		\subfigure[{Full count total link strength} 		\label{fig:full_event_strength}]
		{\includegraphics[height=48mm]{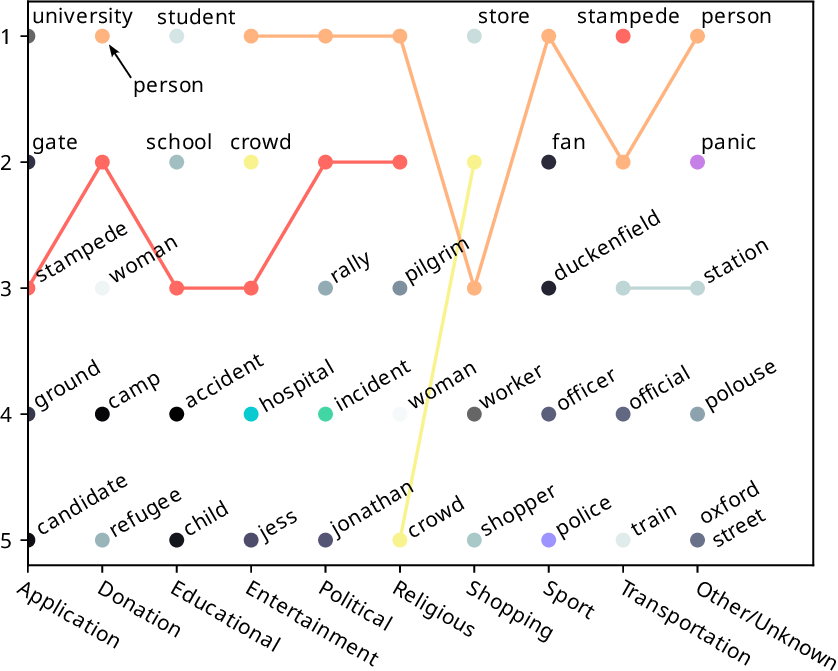}}
    \caption{Words associated with crowd accidents depending on the purpose of gathering. Considering the large number of categories only the top-five words are considered in this analysis.}
		\label{fig:words_event}
\end{figure}

Similar to the results presented earlier when discussing historical and geographical trends, regardless on the purpose of gathering ``stampede'' and ``person'' regularly takes the top-two positions (see \figurename~\ref{fig:words_event}). However, this is mostly true in the binary count, whereas, in the full count, a more variegate image is presented for each type of gathering. For instance words such as ``store,'' ``shopper,'' or ``sale'' appears in accidents related to (bargain) shopping. ``Fan,'' ``stadium,'' and ``match'' are seen in sport events. Further, ``school,'' ``student,'' and ``child'' are often used to describe accidents in educational facilities. But what is most remarkable in the results of \figurename~\ref{fig:words_event} is the use of ``stampede'' in relation to sport events. In fact, it is interesting to note that ``stampede'' plays a less important role in such gatherings. This can be explained considering that accidents in sport events are diverse, with people being crushed at the front of barrier or overwhelmed by a group of people fleeing a violent fight. This particularity of sporting events will be also encountered in the forthcoming sentiment analysis.

\subsection{Sentiment analysis}
So far occurrences (frequencies) and total link strength (co-occurrence) were used to analyze reports on crowd accidents. The lexical analysis already allowed us to highlight some nuances in relation to geographical origin of the reporting institution or gathering nature. However, only sentiment analysis can help in clearly showing emotional differences when reporting crowd accidents. Main findings are discussed here through the use of graphs; numerical results can be found in Appendix C while data used in this work are openly provided in \citep{ZenodoDataset2023B}. Also, it is worth remembering that here only reports published within one year from the accident are considered (still comprehending over 95\% of the total) \footnote{In addition, one report was excluded because it referred to a comment provided by Reuters to describe a picture and had no title.}.

\begin{figure}[htbp]
		\centering
    \includegraphics[width=75mm]{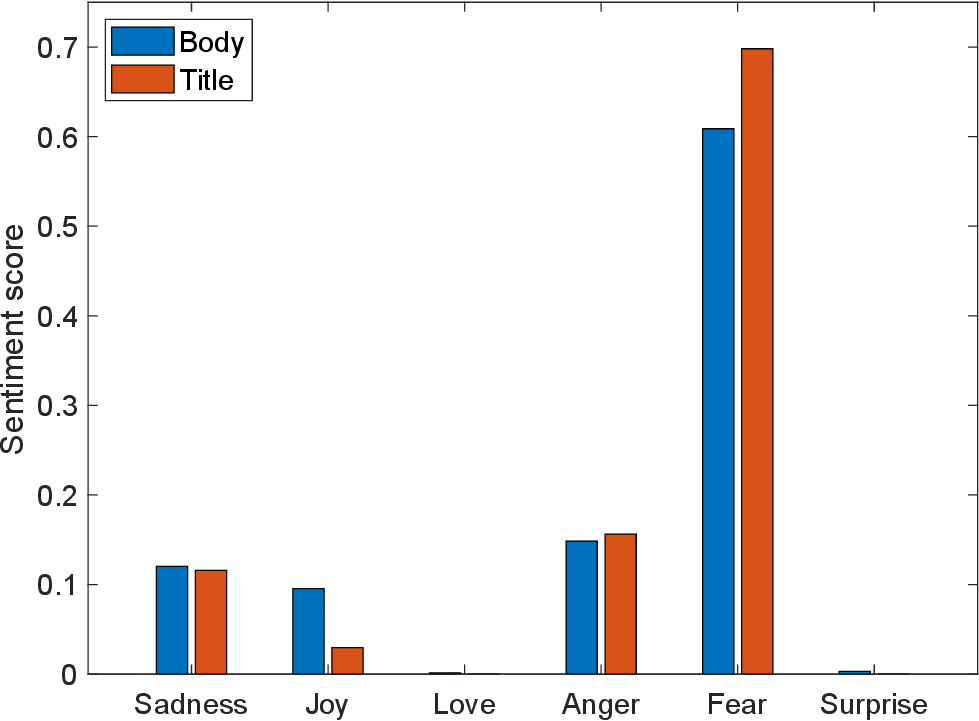}
    \caption{Emotions associated with the complete corpus of press reports on crowd accidents. Emotions were estimated separately using a variant of the DistilBERT model \citep{Sanh2019,DistilBERT,Savani2023} for title and body.}
		\label{fig:sentiment_average}
\end{figure}

We can start the presentation of the sentiment analysis by showing the spectrum of emotions when all reports are considered. Results for the articles' title and body are presented in \figurename~\ref{fig:sentiment_average}. Not surprisingly ``love'' and ``surprise'' are almost completely lacking from the whole corpus of news reports. On the other hand, ``fear,'' ``anger,'' and ``sadness'' are consistently found in both title and body. Interestingly, ``joy'' is also to be found in a significant proportion, especially in the body. This can be explained by the fact that many articles presenting accidents during festivals or joyful events start by describing the atmosphere before the accident to later dig into the dramatic details. This can also explain why ``joy'' is mostly found in the body and not the title which typically focuses only on the dramatic nature of the accident. For these reasons, in the forthcoming analysis, ``love'' and ``surprise'' will be ignored and we will focus mostly on negative emotions plus ``joy.'' \par
The aggregated outcome presented in \figurename~\ref{fig:sentiment_average} is also important to understand the results presented here onward. Crowd accidents are fearful, sad events, and as such, ``fear'' and other negative emotions are expected to be present in most, if not all, reports. Identifying a ``perfect mix'' of emotions for an ideal portrayal is not the scope of this work and is probably an arguable practice. For these reasons, our analysis is of a comparative nature, whereas temporal trends over time are used, and comparisons over specific attributes of crowd accidents are made. Therefore, we advise the reader to go through the results by keeping in mind the comparative nature of our analysis.

\subsubsection{Historical trend}
\begin{figure}[htbp]
		\centering
		\subfigure[{Article body} 	\label{fig:comparison_time_body_full}]
		{\includegraphics[height=50mm]{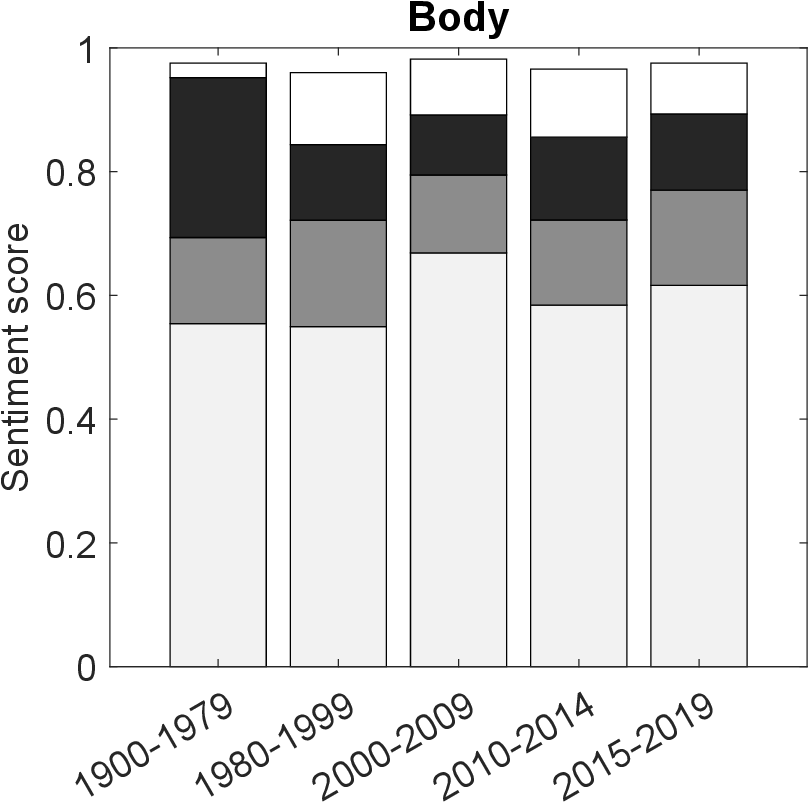}}
		\subfigure[{Article title} 	\label{fig:comparison_time_title}]
		{\includegraphics[height=50mm]{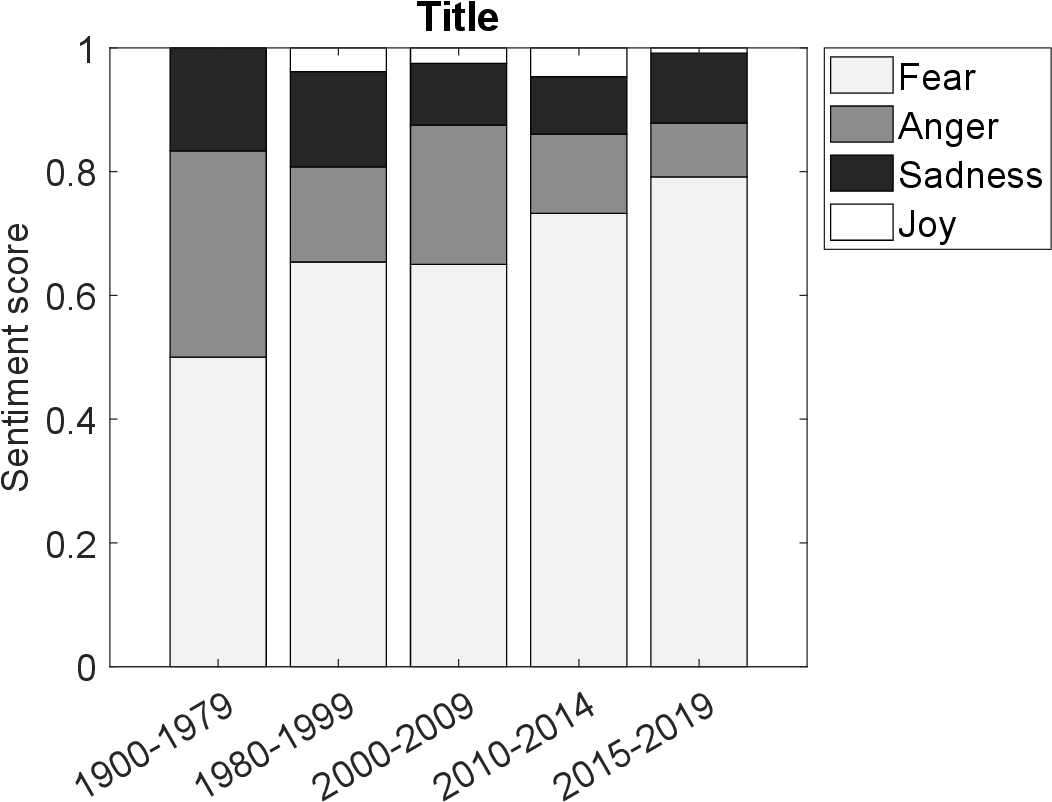}}
    \caption{Historical trend of emotions related to crowd accidents. Total values do not normally reach 1 because a small amount of ``love'' and ``surprise'' is always present.}
		\label{fig:sentiment_time}
\end{figure}

We can now continue the sentiment analysis by following the same order used in the lexical analysis. Historical trends will be considered first, with the evolution in the mix of emotions for title and body presented in \figurename~\ref{fig:sentiment_time}.

As previously mentioned, ``joy'' shows fluctuations possibly related to the specific way in which an accident is introduced, i.e., straight to the point or including a description of the event before the tragedy struck. In the case of the body it is also possible to observe that proportions of ``fear,'' ``anger,'' and ``sadness'' are fairly constant over time, with ``fear'' taking a larger proportion. However, in the title, a clear increase of the ``fear'' emotion is observed over time. ``Fear'' goes from a score of 0.5 in the period from 1900--1979, up to a value close to 0.8 for the more recent period. ``Anger'' shows an opposite but less clear trend: from about 0.35 to slightly less than 0.1. ``Sadness'' is fairly constant over time. This result indicates a tendency to dramatize the events by presenting them in a fearful way in a possible attempt to attract attention from the readers (remember that the trend is seen in the title, which is aimed to attract the attention to an article). It is however worth mentioning that this tendency is not specific to crowd accidents and Rozado et al. \citep{Rozado2022} obtained similar findings by examining a much broader range of news articles. \par
Overall, this result indicates an attempt to leverage on the so-called negativity bias and obtain attention from the readers by presenting events in a negative and scary light. This tendency can hinder the capability of a report to deliver a correct representation of the events. However, by the fact that the negativity bias is mostly used in the title, we can expect that in the body a more-or-less emotionally neutral description can be found.

\subsubsection{Geographical origin}
\begin{figure}[htbp]
		\centering
		\subfigure[{Article body} 	\label{fig:comparison_area_body_full}]
		{\includegraphics[height=50mm]{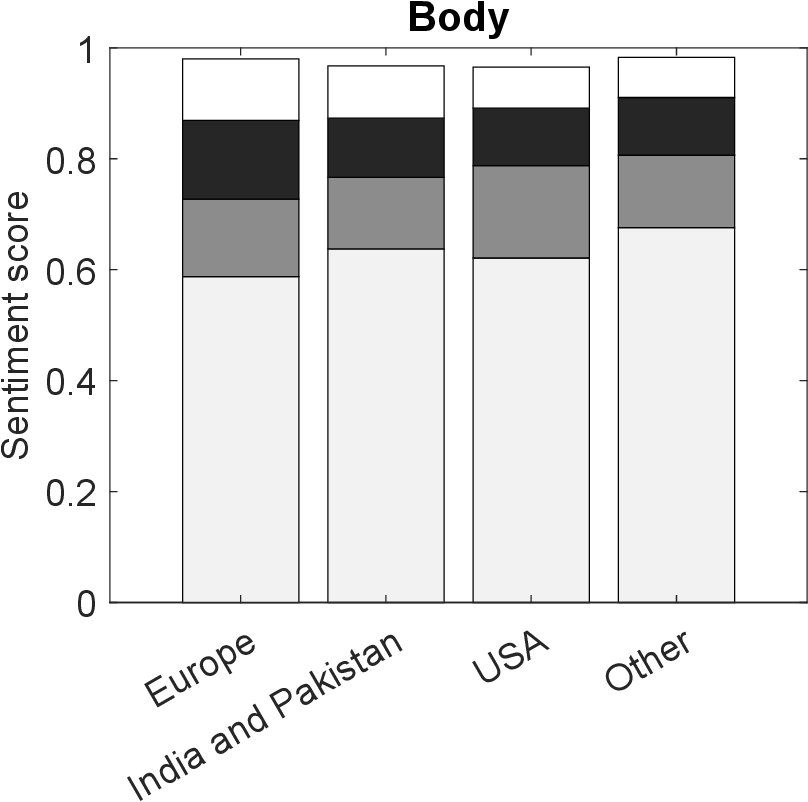}}
		\subfigure[{Article title} 	\label{fig:comparison_area_title}]
		{\includegraphics[height=50mm]{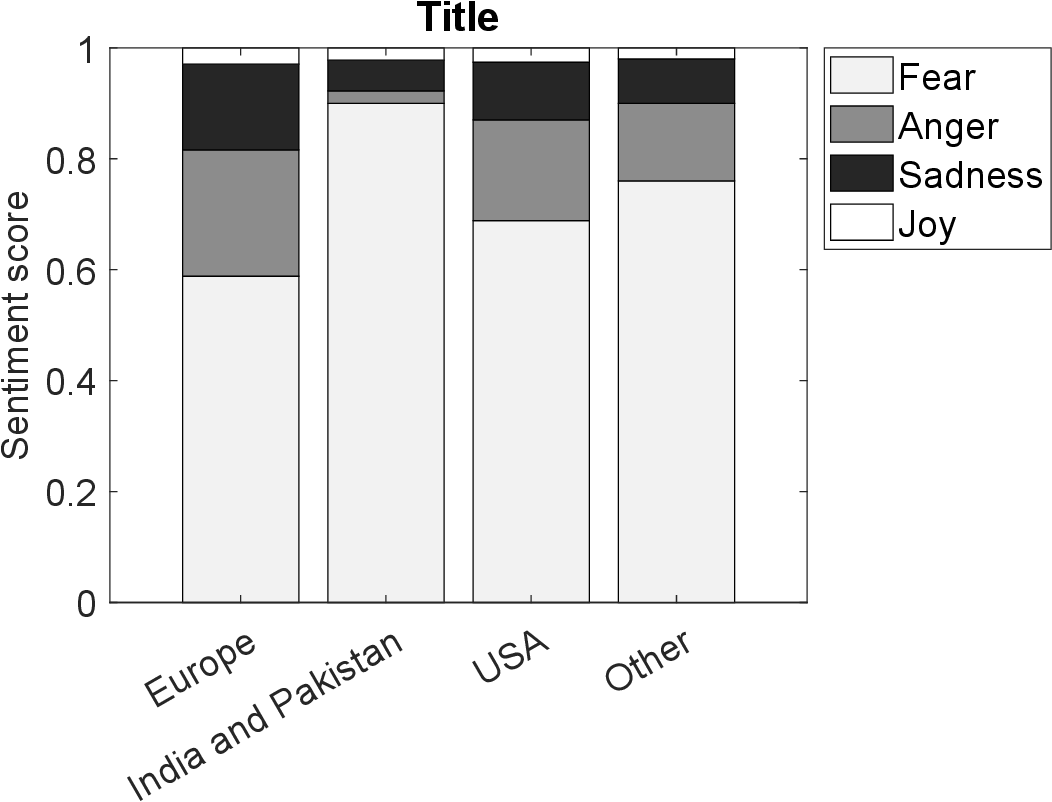}}
    \caption{Sentiments associated to the reporting of crowd accidents in different geographical areas.}
		\label{fig:sentiment_area}
\end{figure}

When the geographical dimension is considered (see \figurename~\ref{fig:sentiment_area}) we can see that, like for historical evolution, differences among areas are only seen in the title.

As already seen for the complete corpus, ``joy'' is also present in small measure in the body and plays a very marginal role in the title. In the body a similar image is obtained for all areas, especially when focusing on the target regions (Europe, India and Pakistan, and the USA). The sentiment of ``fear'' takes similar scores (around 0.60--0.65), with ``sadness'' being slightly higher in Europe and ``anger'' more prominent in the USA. However, the most remarkable feature is seen when title is analyzed. Here, a remarkable amount of ``fear'' is found in news reported from institutions in India and Pakistan (the score equals 0.9). The more common use of the word ``incident'' found in the lexical analysis for India and Pakistan hinted toward a more dramatic representation and the results are now confirmed through the sentiment analysis. However, as we will see, this could partially be linked to the religious nature of most accidents that occurred in India and Pakistan.

\subsubsection{Purpose of gathering}
Finally, we will consider emotions associated with accidents which occurred during different types of gatherings. Considering the large number of categories analyzed here we had to limit our analysis to those having a sufficient number of reports and thus only accidents related to religious, sport, or entertainment events are considered, with results presented in \figurename~\ref{fig:sentiment_event}. The full set of numerical results is reported in Appendix C.

\begin{figure}[htbp]
		\centering
		\subfigure[{Article body} 	\label{fig:comparison_event_body_full}]
		{\includegraphics[width=60mm]{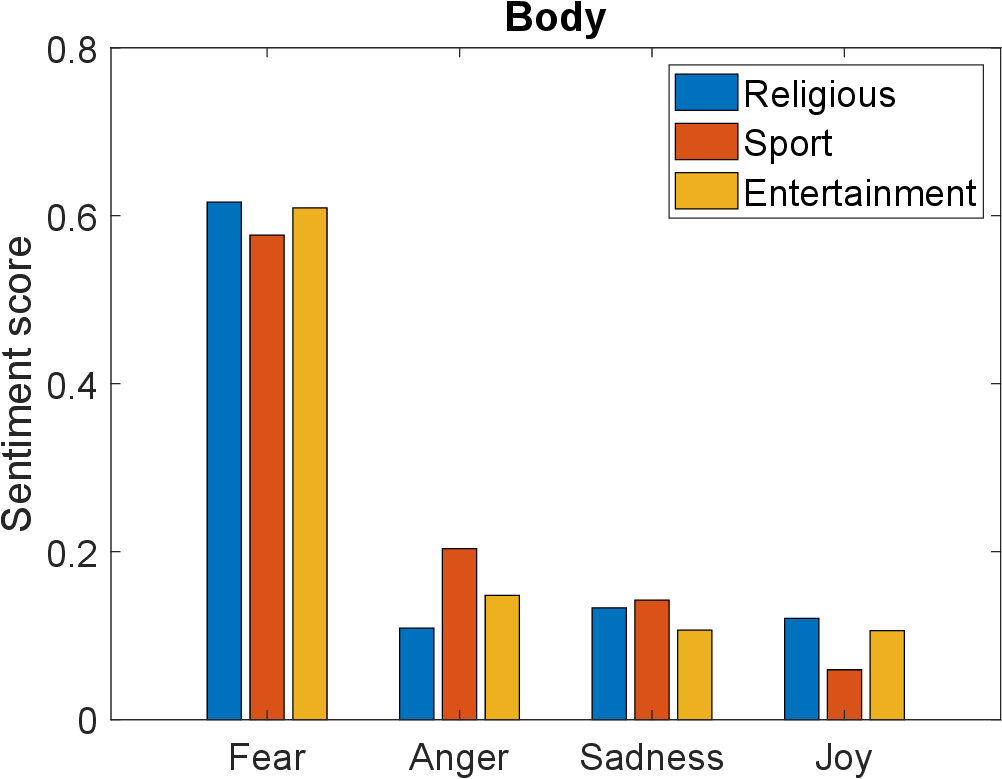}}
		\subfigure[{Article title} 	\label{fig:comparison_event_title}]
		{\includegraphics[width=60mm]{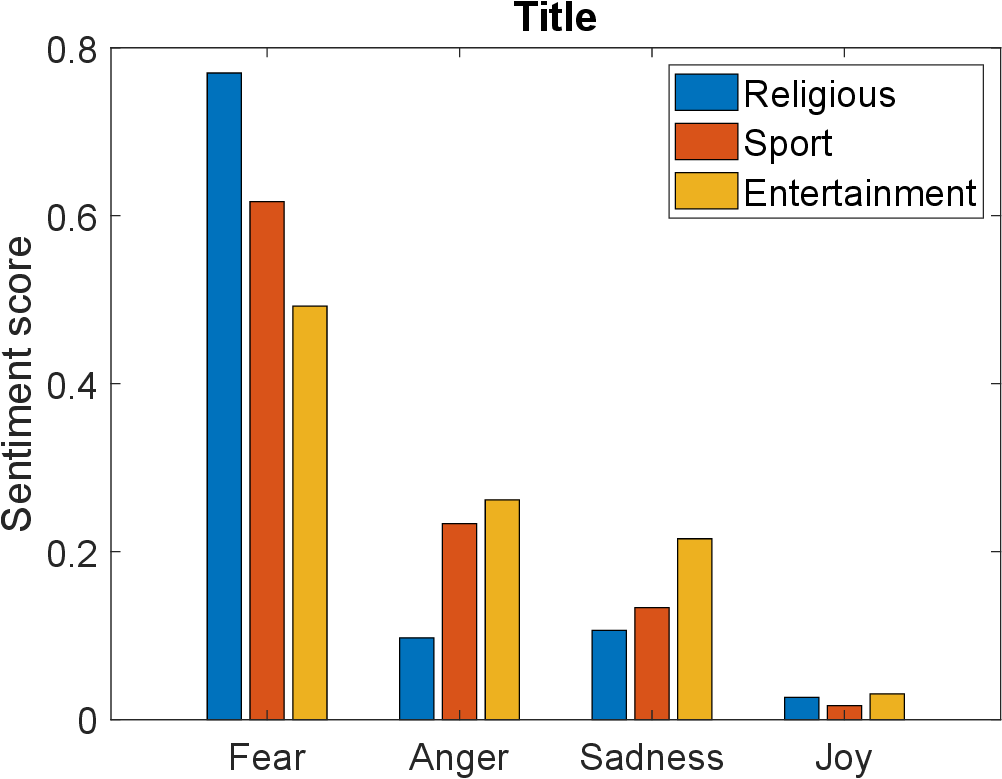}}
    \caption{Share of emotions in reports focusing on different types of events. Only gatherings' categories with a sufficient number of reports (over 50) were selected; accidents related, for example, to application or shopping have less than 10 reports making a valid analysis difficult.}
		\label{fig:sentiment_event}
\end{figure}

Even from this perspective differences among events appear clear only in the title. Specifically, it is suggested that religious events carry a more dramatic dimension, with higher levels of ``fear'' compared to sport and entertainment. This result may be linked with what previously found for India and Pakistan where religious events are held in large numbers and result in a considerable amount of accidents. However it is not clear whether the nature of events influence the geographical result or the way around, only the link can be observed. \par
The higher scores of ``anger'' and ``sadness'' in the title of sport and entertainment accidents can be possibly explained with the presence of organizers and, in the case of soccer, of a rival fan club. Religious events can lead to a large number of fatalities and the crowd is usually managed by a large number of stakeholders mostly unknown by the attendants and may also include volunteers. Therefore it is harder to blame someone specific and accidents tend to be represented as a scary tragedy in which innocent people lost their lives. But, especially in soccer matches, stadium staff or security personnel are often targeted as responsible or fanclubs blame each other for the misconduct. Under such circumstances we can expect angry headlines hinting at specific individuals or institutions as responsible and calling the tragedy predictable or voluntarily caused.

\subsection{Wikipedia before and after Itaewon}
On the night of October 29th, 2022 a crowd accident claiming 159 people's lives occurred in the neighborhood of Itaewon in Seoul, South Korea \citep{WikiItaewon}. The magnitude of the accident, the fact that it occurred in a major country playing at the world stage and the circumstances under which the accident occurred, all contributed to creating a lot of media attention on the topic of crowd management and safety. Media outlets all over the world covered the accident and practitioners and academics have been often invited to contribute in press articles or act as experts in TV and radio programs. \par
Although crowd accidents occur almost on a monthly basis \citep{Feliciani2023}, a small portion get media attention for a relatively prolonged period of time. Itaewon was probably one of those and even people not involved with crowds in the frame of their work or research were discussing the causes of such tragedies and how to prevent them. In such a context, many academics and experts had the task to mitigate the myths associated with crowd disasters and tried to promote the use of correct words to describe phenomena and mechanisms observed in such circumstances. \par
It may be therefore relevant to verify whether the media coverage from such a tragedy and the involvement of experts in the analysis had an influence on the terms used in describing crowd accidents. For this purpose we decided to compare the very same list of crowd accidents listed in Wikipedia and check whether description had been modified after the tragedy in Itaewon and how it has changed. By chance the content of Wikipedia pages had been programmatically scraped on October 15th, 2022 (i.e., two weeks before the accident). The same pages were later consulted on May 25th, 2023 and the content compared to recognize any modification. We decided to not focus on the press for this comparative analysis because, as discussed earlier, long time periods and a large number of reports are needed to see clear trends. On the other hand, Wikipedia pages can get modified over time and contributors may want to add material and revise content that was found being incorrect. For this reason even small changes over a short time period are easy to recognize and track.

\begin{figure}[htbp]
		\centering
		\subfigure[{Decreasingly used} 	\label{fig:wiki_down}]
		{\includegraphics[width=60mm]{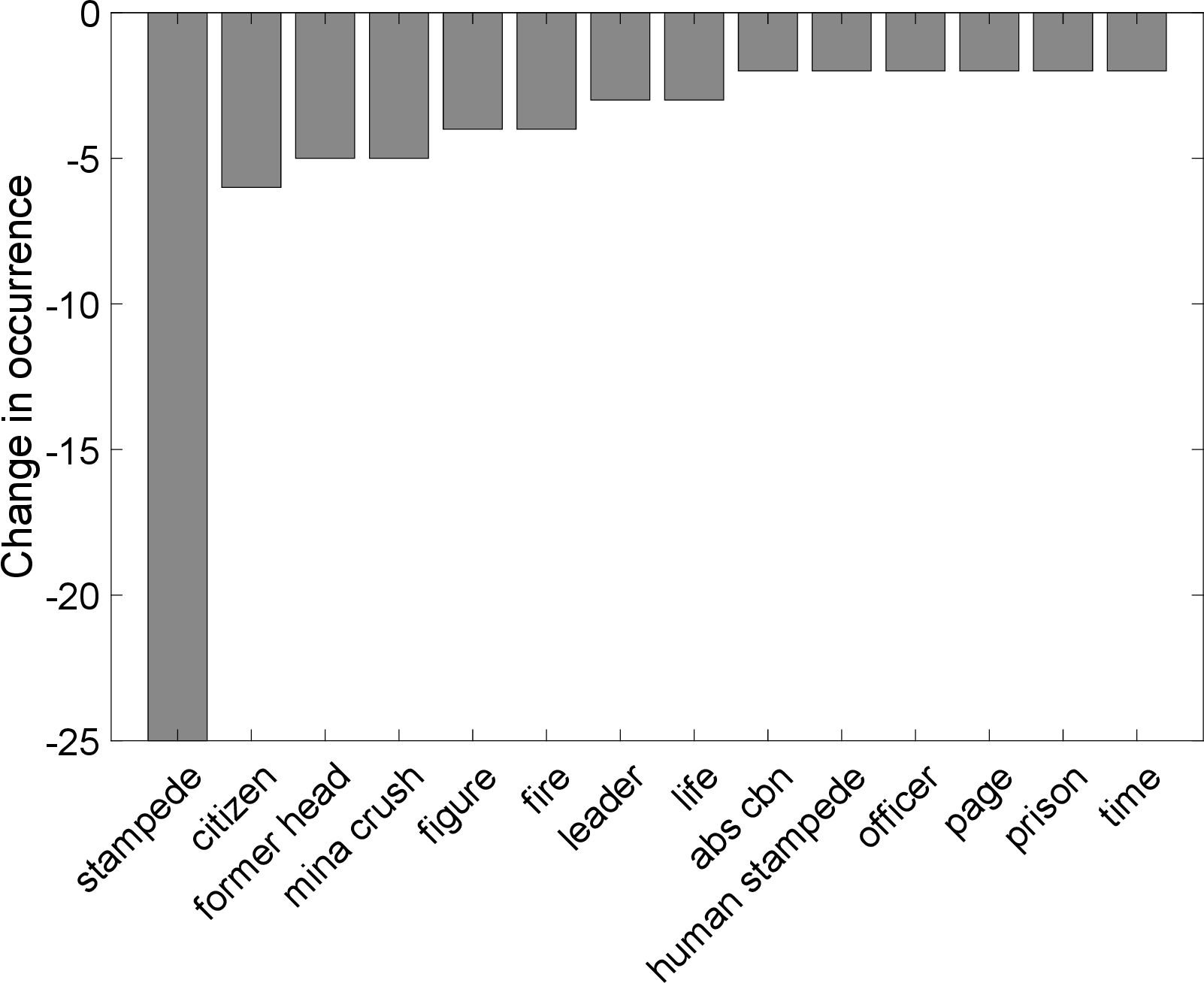}}
		\subfigure[{Increasingly used} 	\label{fig:wiki_up}]
		{\includegraphics[width=60mm]{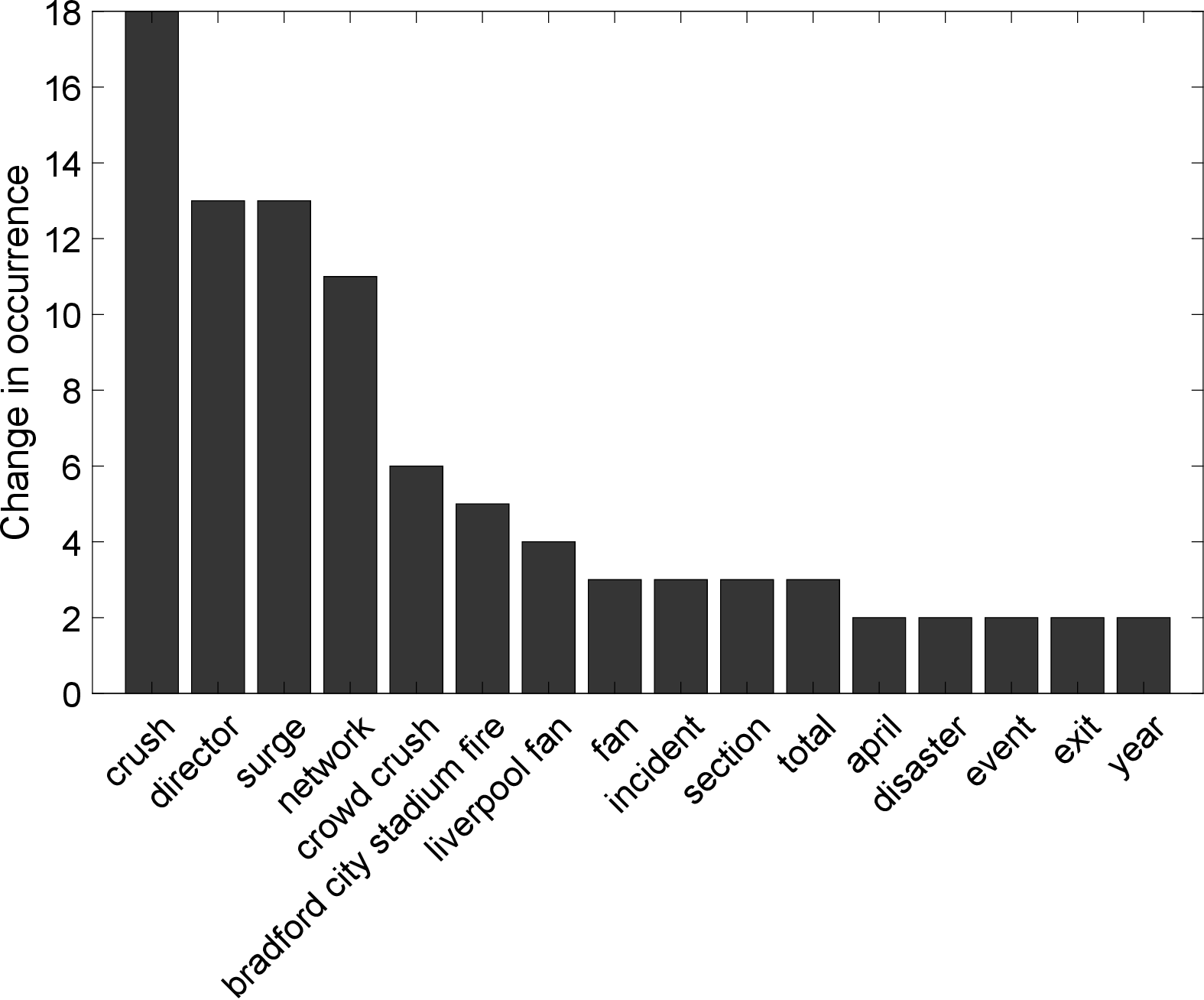}}
    \caption{Change in the occurrences of words in Wikipedia pages presenting crowd accidents following the Itaewon tragedy. Texts were preprocessed using VOSviewer to remove articles or prepositions and perform plural to singular conversions (among others). Only words appearing at least five times were considered in the analysis. Title and body were combined in performing the analysis.}
		\label{fig:wiki_results}
\end{figure}

Results for this analysis are presented in \figurename~\ref{fig:wiki_results}. It is remarkable to see that ``stampede'' has been the word that was most frequently removed from Wikipedia pages following the accident in Itaewon. On the other hand, the use of ``crush'' has increased (especially if ``crush'' and ``crowd crush'' are considered together). Although this result can already imply a trend in replacing ``stampede'' with ``crush,'' the results of \figurename~\ref{fig:wiki_results} do not necessarily prove this hypothesis. An ad hoc analysis was therefore subsequently performed by checking more specifically how texts were modified. Results (partially presented in \tablename~\ref{tab:wiki_changes}) show that ``stampede'' was changed into ``crush'' in ten instances and into ``surge'' in three instances. Although the changes are minimal in a corpus of over 65,000 words it is worth noting that ``stampede'' clearly stands out as a word deserving attention, with additional modifications being limited to irrelevant changes.

\begin{table}[htbp]
  \caption{Modifications produced to the 68 individual pages listing crowd accidents in Wikipedia during the six-months period following the tragedy of Itaewon. Only instances larger or equal two are reported in the table. In this analysis, raw text data are used (i.e., no pre-processing using VOSviewer), which explains the presence of ``the,'' ``in,'' etc.}
	\label{tab:wiki_changes}
\begin{tabular}{lll}
  \hline
	Original word 	& Modified word		& Instances	\\
  \hline
	stampede(s)			& crush(es)				& 10				\\
	stampede				& surge						& 3					\\
	Chief						& chief						& 2					\\
	Van							& van							& 2					\\
	head						& director				& 2					\\
	in							& with						& 2					\\
	this						& the							& 2					\\
	\hline
\end{tabular}
\end{table}

More generally, the analysis of the Wikipedia corpus shows that major catastrophic events followed by broad media coverage of an incident that is particularly founded in experts' input has the potential to trigger a revision of terms used to describe crowd accidents, although we cannot confirm a casual relationship between the modifications and the Itaewon accident. What can be more certainly confirmed is that the word ``stampede'' is gradually being reconsidered and ``crush'' and ``surge'' are gaining popularity as terms in relation to crowd accidents.

\section{Discussion}
In presenting the results we focused the discussion on specific lexical and sentimental aspects relative to each outcome, mostly describing each graph individually. Here, we want to combine the multiple findings and discuss what they imply in regard to the reporting of crowd accidents. \par
Our analysis suggests that among the controversial words usually associated with crowd behavior, only ``stampede'' can be consistently found in press reports. On the other hand, ``panic,'' which is arguably the most controversial, did not take a central role in the popular press although it was consistently found in scientific literature on crowds. However, we should remark that although ``panic'' did not typically make the top-ten in the rankings (with a few exceptions, see \figurename~\ref{fig:words_time}, \figurename~\ref{fig:words_area}, and \figurename~\ref{fig:words_event}), it was nevertheless found in lower positions in most analyses (e.g., it takes position 13 for the period 2015--2019 when considering total link strength and using full count; see Appendix D for more details). Since the totality of the press reports considered here covers crowd accidents, and panic behavior is reportedly rare in such occasions, it is therefore problematic if this word consistently appears at all, regardless of how common it is. Also, we should mention that ``stampede'' already partially contains the notion of panic and therefore its presence already hints to a description in which people are portrayed as behaving irrationally and where chaos reigns. Furthermore, when it comes to the use of ``panic'' in scientific publications, we cannot exclude that some works employ this word to criticize its improper use, as has been done in this work. \par
However, we should also acknowledge the limitations regarding the lexical analysis, especially when it comes to the context. As already briefly outlined, we cannot rule out that controversial words are sometimes used in a critical way. In addition, it is also necessary to consider that, especially in the scientific literature where many are not native English speakers, ``panic'' and ``stampede'' are used as synonyms for non-controversial words without overly negative connotations intended. Often, ``panic'' is (mistakenly) used as a synonym for urgency (see, for example, \citep{Chen2023}), while ``stampede'' is also used to simply refer to crowd accidents (e.g., \citep{Yi2020}) or merely inspired by the work of influential authors \citep{Helbing2000}. Such inappropriate use should be considered differently from native English-speaking (professional) journalists. Nonetheless, the increasing use of controversial words becomes problematic when the target meaning deviates from the dictionary definition or public understanding. In this sense, when discussing crowd accidents, writers should be aware of controversial words and check reliable resources (dictionaries, etc.) to avoid the unwanted manipulation of the public perception on crowd behavior and phenomena. \par
As discussed above, our analysis suggests that there is a problem in relation to the choice of words in crowd accidents reports. However, this problem is not simply due to a few words (like ``panic'' or ``stampede''), but a more general issue arising from a misleading overall representation, as complemented by the sentiment analysis. In other words, when the results are taken altogether, we see that there is an increasing tendency in dramatizing crowd accidents by picking up words such as ``stampede'' to attract the attention of readers. However, the opposite mechanism, i.e., unintentional word choice leading to more dramatic representations, is also possible. Regardless of the causal relationship, we highlighted a general tendency toward fearful representations of crowd accidents with subtle differences given by cultural background and the nature of the accident. \par
However, we also need to acknowledge limitations with the sentiment analysis. For instance, ``anger'' can be regarded as a negative sentiment, but there are cases in which such emotion is not necessarily related to a misleading portrayal of crowd accidents. If some stakeholder (e.g., organizer, fan club, police, etc.) is found negligent, anger is a predictable reaction from the victims and can lead to better safety measures in the future. In this sense, we should remind the readers once again that our approach is comparative and relative to the whole corpus of text. As such, care is always needed when considering specific cases. \par
To conclude this discussion, we should also note that although our analysis hints at an increasingly problematic representation of crowd accidents; the case of Itaewon and the modifications in Wikipedia also suggest that in some contexts \footnote{Wikipedia can be considered a popular press source, but it relies on different mechanisms compared to traditional media outlets.}, a partially opposite trend is observed, with more neutral and accurate descriptions. In summary, long-term trends suggest a tendency toward dramatic descriptions. However, it is possible that this trend is being reversed in favor of factual reporting if those writing crowd accident reports spend more time seeking accurate representations and collaborate with experts in the field.

\section{Conclusions}
In this work, reporting of crowd accidents in popular media has been investigated through lexical and sentiment analysis. Lexical analysis allowed us to highlight the most commonly used and most relevant words in crowd accident reports, with ``person'' (or its plural ``people'') and ``stampede'' being constantly among the top-two. Differences in relation to the geographical area of reporting or the purpose of gathering were also highlighted, although these typically concerned less relevant words such as ``incident,'' ``crowd,'' ``death,'' or ``police.'' On the other hand, sentiment analysis showed a trend toward the dramatization of crowd accidents. This effect, although general, appears to be slightly stronger in India and Pakistan and/or during religious events, although the causal link between these elements remains unclear. \par
Our results suggest that a more neutral representation of crowd accidents is needed with the use of proper words being only the beginning of this process. The gradual change from ``stampede'' into ``crush,'' although limited in number or context, and arguable in choice, brings hopes toward better reports on crowd accidents. In an effort to address professionals involved in news reporting, we want to highlight that although choosing a technically correct word (e.g., ``crush,'' ``surge,'' ``trampling,'' ``progressive crowd collapse,'' etc.) is certainly important to provide accurate descriptions, a minimalistic neutral description simply announcing an accident in general terms (e.g., ``a crowd accident occurred'') can have an informative nature and yet allow to gain time while waiting for details and consult with experts for a more detailed coverage. In other words, providing answers to the five Ws (i.e., who, what, when, where, and why) ensures a neutral yet comprehensive coverage of an accident. Specifically, ``what, when, and where'' should be clarified from the outset as these are undisputed facts. On the other hand, determining ``who and why'' may require additional time, and in such cases, consulting with professionals experienced in crowd dynamics is advisable to ensure accurate descriptions of the events. \par
While journalists play a crucial role in reporting tragic events and have the potential to drive a shift towards a more nuanced portrayal of crowd accidents, we acknowledge that changes in reporting styles can be challenging to enforce. Moreover, they are often intertwined with rapidly evolving social norms, propagated through social networks and other online communities. In this context, our work serves as an initial effort to raise awareness within the research community about the significance of word choice. We hope that this awareness will catalyze better practices not only in journalism but also in various other domains. \par
In addition to the implications for crowd and safety research, we should also stress on the novelty of our work. The combination of lexical and sentiment analysis is certainly new for the case of crowd accidents. However, to the best of our knowledge, it may also represent a novel approach from a wider perspective. While showing specific features in regard to the representation of crowd accidents, our combined approach also showed its potential by highlighting a more complete picture on the reporting style compared to what could have been achieved by employing only a single technique. \par
To conclude, we believe our work could help improve the awareness and preparedness against crowd accidents while also introducing a methodological approach which could find applications in other areas of research. Future research could extend the time frame of our analysis while also exploring more complex inter-correlations. For instance, investigating whether the dramatic reporting in India and Pakistan exhibits cultural traits or is linked to the religious nature of mass gatherings would provide valuable insights. Additionally, exploring the potential relationship between dramatization and the geographical proximity of an accident could broaden the scope of this work.

\section*{Acknowledgments}
This work was financially supported by the JST-Mirai Program Grant Number JPMJMI20D1 and the JSPS KAKENHI Grant Number JP23K13521. In addition, the authors would like to express their gratitude to Pietro Beretta Piccoli for his help in extracting texts from online media outlets. Finally, we would like to express our gratitude to the two anonymous reviewers for their valuable feedback.

\section*{Appendix A}
This appendix details the method and the process followed to create a corpus of press reports on crowd accidents. Presentation will follow by explaining the list of crowd accidents on which this work is partially based to later detail on the choice of sources and data extracted from each report.

\subsection*{Creation of a list on past crowd accidents}
The first step to build a dataset containing representative media reports on crowd accidents is to define what is a crowd accident and create a list of past events. This work was already carried out in \citep{Feliciani2023} where 281 crowd accidents were identified for the period between 1900 and 2019. Here, to focus the discussion on aspects relevant to this work we will briefly summarize characteristics of the list and address readers interested in details to the original work. \par
The working definition of crowd accident is an important element to understand the outcome from this work and can be summarized as follows. A crowd accident is considered as such when one person or more than 10 were injured in a crowded context. We consider as a crowd accident only those in which crowd motion can be considered the main cause for fatalities and thus excluded, for example, those cases in which fire and smoke intoxication played a major role or when most or all victims were killed through the use of weapons. In short, we consider those cases that are commonly (although often mistakenly) labeled as ``stampede'' or ``crush.''

\subsection*{Selection of representative media organization}
Once a list with crowd accidents is prepared, media reports need to be collected. This task comes with a number of challenges, which, when properly tackled, can guarantee that collected material reflects a valid sample of press reports. What is important to remark is that collecting all reports from all media outlets for all accidents is certainly impossible and also likely inaccurate. Many media outlets rely on news agencies (AFP, AP, Reuters, etc.) and thus reports from different sources are almost identical. Also, although blog or official pages from fanclubs can provide reliable and important information on accidents which occurred in soccer events, writers are usually lay people with a different background from professional journalists. We therefore had to define a strategy which would help to: 1) limit reports from reputed sources, 2) avoid redundant reports largely based on news agencies, 3) ensure that reports reflect reporting style of their age and geographical location.

\begin{figure}[htbp]
		\centering
    \includegraphics[width=110mm]{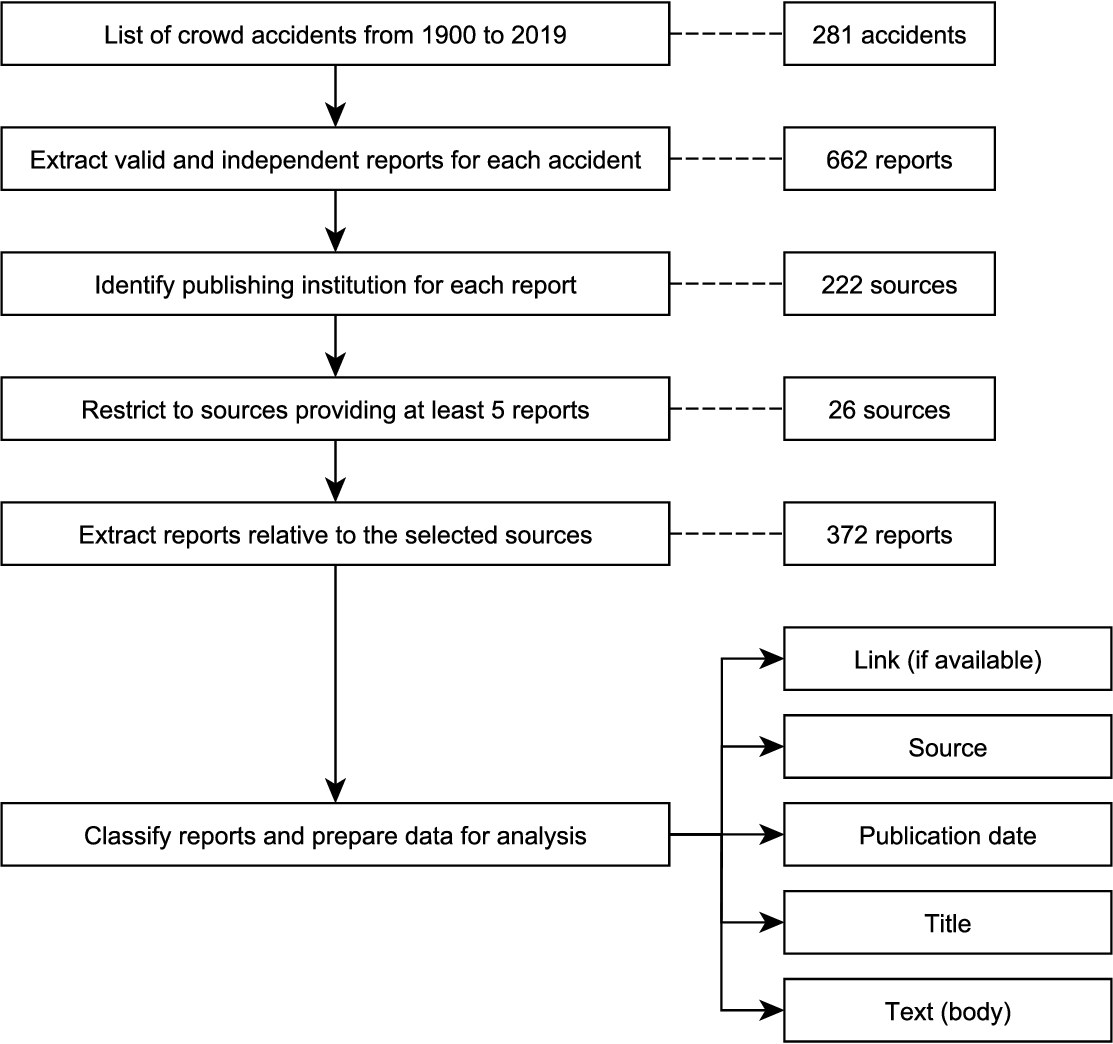}
    \caption{Schematic representation of the method employed to select and classify media reports for crowd accidents. The approach taken to generate the initial list of 281 accidents is described in \citep{Feliciani2023} and the list is provided in \citep{ZenodoDataset2023A}.}
		\label{fig:report_selection}
\end{figure}

In light of the above considerations, we used the approach schematically presented in \figurename~\ref{fig:report_selection} to select representative reports. The approach can be summarized as follows. For each accident a few reports were extracted based on internet search or by looking into newspaper archives for older events. We tried to collect about 2--3 reports for each accident by making sure sources were as independent as possible. For instance, reports from different outlets but based on text from the same news agency were avoided. Each report went through a simple check to verify content is reliable and can be generally trusted. After this initial process, a list containing all sources was created. From a total of 662 reports 222 independent sources were identified. This comparatively large number can be explained when considering that some accidents are reported by local news outlets which are not likely to report small accidents occurring in other parts of the world. In addition, fanclubs are often a precious source of information for soccer-related events, but usually only cover tragedies related to a specific club. It should be noted that this initial list also included reports in languages other than English because its additional purpose was to collect numerical information on fatalities, injuries, and crowd size to be used in a different work \citep{Feliciani2023}. To consider only sources having a global or at least a regional coverage we consequently decided to focus only on those media outlets which covered at least 5 accidents listed in our dataset. These media outlets are listed in \tablename~\ref{tab:source}.

\begin{table}[htbp]
  \caption{News outlets used to construct the final dataset of media reports. ``Type'' is used to roughly indicate the traditional reporting style for each organization. Some have multiple formats and may not exist anymore in paper. ``Reports'' indicate the number of entry used in our dataset.}
	\label{tab:source}
\begin{tabular}{lllll}
  \hline
	Organization   			& Country        	& Type         	& Founded		& Reports 	\\
  \hline
	BBC                	& United Kingdom 	& Broadcaster   & 1922    	&	78				\\
	New York Times     	& USA            	& Newspaper     & 1851    	&	31				\\
	The Guardian       	& United Kingdom 	& Newspaper     & 1821    	&	30				\\
	The Times of India 	& India          	& Newspaper     & 1838    	&	22				\\
	CNN                	& USA            	& Broadcaster   & 1980    	&	21				\\
	Reuters            	& United Kingdom 	& News agency   & 1851    	&	20				\\
	Al Jazeera         	& Qatar          	& Broadcaster   & 1996    	&	17				\\
	Hindustan Times    	& India          	& Newspaper     & 1924    	&	17				\\
	India Today        	& India          	& News magazine & 1975    	&	17				\\
	China Daily        	& China          	& Broadsheet    & 1981    	&	16				\\
	AP                 	& USA            	& News agency   & 1846    	&	11				\\
	The Independent    	& United Kingdom 	& Broadsheet    & 1986    	&	10				\\
	ABC                	& Australia      	& Broadcaster   & 1932    	&	8					\\
	France 24          	& France         	& Broadcaster   & 2006    	&	8					\\
	UPI                	& USA            	& News agency   & 1907    	&	7					\\
	NBC                	& USA            	& Broadcaster   & 1926    	&	6					\\
	News24             	& South Africa   	& Online news   & 1998    	&	6					\\
	The Indian Express 	& India          	& Newspaper     & 1932    	&	6					\\
	The Tribune        	& India          	& Newspaper     & 1881    	&	6					\\
	Business Standard  	& India          	& Newspaper     & 1975    	&	5					\\
	Dawn               	& Pakistan       	& Newspaper     & 1941    	&	5					\\
	Fox News           	& USA            	& Broadcaster   & 1996    	&	5					\\
	IOL                	& South Africa   	& Online news   & N/A     	&	5					\\
	NDTV               	& India          	& Broadcaster   & 1984    	&	5					\\
	The Hindu          	& India          	& Newspaper     & 1878    	&	5					\\
	oneindia           	& India          	& Online news   & 2006   		&	5					\\
	\hline
\end{tabular}
\end{table}

As shown in \tablename~\ref{tab:source} newspapers and broadcasters from the UK and USA were among the most common sources, although Indian media are also generally well represented. The strong presence of English media is attributed to the use of English as the primary working language. While other languages were also utilized, ``local'' languages were predominantly employed for reporting on local events, resulting in their under-representation in the dataset. For instance, we found a useful Indonesian article from a local media outlet for understanding the crowd size in a related accident in Indonesia. However, no additional reports from that media outlet were included, preventing it from meeting the minimum requirement of five reports for inclusion in the final dataset used for analysis. In short, only original English reports are used in our analysis. From the list of \tablename~\ref{tab:source}, it can be also seen that old organizations tend to be represented with larger numbers of reports. This is an indirect consequence of the approach taken when selecting the reports and could be seen as a partial proof of its consistency: old organizations have a longer history and tradition and can therefore represent changes in reporting style and information technology.

\subsection*{Extracted information and categorization}
Following the approach illustrated above and presented in \figurename~\ref{fig:report_selection}, a total of 372 reports were selected for the final dataset and were used in the analysis. For each report the following information was extracted.

\begin{itemize}
	\item \textbf{Link:} This allows us to retrieve the text of the report and access to the original article. Due to copyright restrictions the full dataset cannot be shared, but readers interested in reproducing it shall use the link to get access to the content.
	\item \textbf{Source:} The organization publishing the report. This is one among the 26 organizations provided in \tablename~\ref{tab:source}.
	\item \textbf{Publication date:} The date in which the report was published. This is important because some reports describe accidents which occurred several years before, for example when a trial is ongoing and a decision is taken. The reporting style is therefore more closely related to the publishing date rather than the year when the accident occurred.
	\item \textbf{Title:} This is what is also typically known as the headline. Byline, lead or subtitle are excluded from the dataset to increase consistency between the sources.
	\item \textbf{Text:} This is what is normally referred as the body. It contains the description of the accident and all related information. Captions for pictures or other pieces of information relevant for the understanding but not strictly part of the body are excluded.
\end{itemize}

As already briefly stated, the whole dataset cannot be shared due to copyright regulations. However, a list with sources and minimal information allowing an independent recreation of the dataset is openly available at \citep{ZenodoDataset2023B}.

\section*{Appendix B}
This appendix contains the full results from the lexical analysis including scores associated with occurrences and total link strength. Tables are divided into historical trends, geographical origin, and purpose of gathering to reflect the presentation in the main text.

\subsection*{Historical trend}
In the tables provided below time periods are provided horizontally (like for the graphs in the main text) and ranking follows a vertical order.
\begin{table}[htbp]
	\small
	\caption{Words used in crowd accidents reports ranked based on the binary count approach and by considering occurrences within each time interval. This table is equivalent to the representation of \figurename~\ref{fig:binary_time_occurrence}.}
	\label{tab:time_binary_occurrences}
	\begin{tabular}{llllll}
		\hline
			 & 1900--1979   & 1980--1999     & 2000--2009     & 2010--2014    & 2015--2019    \\
		\hline
		1  & person (10)  & person (37)    & stampede (125) & stampede (81) & stampede (93) \\
		2  & crowd (9)    & stampede (33)  & person (117)   & person (76)   & person (86)   \\
		3  & day (9)      & crowd (27)     & hospital (82)  & incident (49) & hospital (54) \\
		4  & body (8)     & tragedy (23)   & official (66)  & police (43)   & incident (47) \\
		5  & disaster (8) & hospital (22)  & death (61)     & crowd (42)    & police (45)   \\
		6  & crush (7)    & death (21)     & crowd (56)     & woman (40)    & death (37)    \\
		7  & death (7)    & year (21)      & incident (55)  & thousand (39) & crowd (36)    \\
		8  & police (7)   & crush (20)     & woman (54)     & hospital (37) & thousand (36) \\
		9  & way (7)      & incident (19)  & police (51)    & official (37) & year (35)     \\
		10 & panic (6)    & authority (18) & city (47)      & death (34)    & place (33)  	\\
		\hline
	\end{tabular}
\end{table}
\begin{table}[htbp]
	\small
	\caption{Words used in crowd accidents reports ranked based on the binary count approach and by considering total link strength within each time interval. This table is equivalent to the representation of \figurename~\ref{fig:binary_time_strength}.}
	\label{tab:time_binary_strength}
	\begin{tabular}{llllll}
		\hline
				& 1900--1979   	 & 1980--1999      & 2000--2009      & 2010--2014      & 2015--2019      \\
		\hline
		1 	& person (124)   & person (1075)   & stampede (3358) & stampede (2210) & stampede (2321) \\
		2 	& day (122)      & crowd (861)     & person (3334)   & person (2158)   & person (2196)   \\
		3 	& disaster (112) & stampede (839)  & hospital (2468) & incident (1410) & hospital (1334) \\
		4 	& crowd (110)    & tragedy (775)   & official (2005) & crowd (1321)    & incident (1308) \\
		5 	& body (109)     & year (754)      & death (1899)    & police (1294)   & police (1215)   \\
		6 	& way (101)      & death (736)     & crowd (1887)    & woman (1213)    & death (1065)    \\
		7 	& death (99)     & hospital (690)  & incident (1527) & thousand (1202) & year (1058)     \\
		8 	& police (99)    & authority (655) & police (1524)   & hospital (1110) & crowd (1013)    \\
		9 	& crush (91)     & time (648)      & woman (1477)    & death (1104)    & place (975)     \\
		10 	& tragedy (89)   & police (646)    & city (1407)     & official (1065) & thousand (969)  \\
		\hline
	\end{tabular}
\end{table}
\begin{table}[htbp]
	\small
	\caption{Words used in crowd accidents reports ranked based on the full count approach and by considering occurrences within each time interval. This table is equivalent to the representation of \figurename~\ref{fig:full_time_occurrence}.}
	\label{tab:time_full_occurrences}
	\begin{tabular}{llllll}
		\hline
				& 1900--1979    & 1980--1999    & 2000--2009     & 2010--2014     & 2015--2019     \\
		\hline
		1 	& person (53)   & person (212)  & person (441)   & person (349)   & person (364)   \\
		2 	& fan (41)      & stampede (81) & stampede (437) & stampede (313) & stampede (307) \\
		3 	& disaster (32) & fan (78)      & hospital (141) & police (94)    & police (99)    \\
		4 	& crowd (30)    & year (71)     & woman (139)    & incident (88)  & incident (93)  \\
		5 	& game (27)     & crowd (61)    & official (126) & crowd (87)     & hospital (86)  \\
		6 	& child (25)    & police (61)   & crowd (115)    & hospital (65)  & stadium (79)   \\
		7 	& year (25)     & stadium (60)  & student (105)  & woman (64)     & official (62)  \\
		8 	& body (23)     & game (50)     & police (96)    & stadium (60)   & woman (61)     \\
		9 	& ground (23)   & time (50)     & incident (94)  & devotee (55)   & year (61)      \\
		10 	& time (21)     & death (48)    & stadium (93)   & thousand (54)  & crowd (54)     \\
		\hline
	\end{tabular}
\end{table}
\begin{table}[htbp]
	\small
	\caption{Words used in crowd accidents reports ranked based on the full count approach and by considering total link strength within each time interval. This table is equivalent to the representation of \figurename~\ref{fig:full_time_strength}.}
	\label{tab:time_full_strength}
	\begin{tabular}{llllll}
		\hline
				& 1900--1979      & 1980--1999          & 2000--2009       & 2010--2014       & 2015--2019       \\
		\hline
		1 	& fan (8518)      & person (49564)      & person (31228)   & person (23180)   & person (22451)   \\
		2 	& stairway (6291) & duckenfield (32431) & stampede (22735) & stampede (18271) & stampede (16120) \\
		3 	& ranger (6205)   & officer (29924)     & crowd (12400)    & police (6813)    & police (6236)    \\
		4 	& person (5828)   & police (23728)      & hospital (9248)  & crowd (6156)     & incident (6089)  \\
		5 	& crowd (5791)    & family (20325)      & official (7765)  & incident (6025)  & hospital (4755)  \\
		6 	& year (5458)     & year (19540)        & woman (7387)     & hospital (4664)  & year (4375)      \\
		7 	& game (5187)     & force (18769)       & police (6409)    & fan (4596)       & official (4274)  \\
		8 	& man (5184)      & evidence (17163)    & death (6333)     & woman (4040)     & stadium (3618)   \\
		9 	& time (5077)     & day (16740)         & fan (6149)       & year (3879)      & death (3369)     \\
		10 	& stair (4979)    & wright (16240)      & child (6134)     & stadium (3728)   & station (3365)   \\
		\hline
	\end{tabular}
\end{table}

\subsection*{Geographical origin}
In the tables provided below geographical regions are provided horizontally (like for the graphs in the main text) and ranking follows a vertical order.
\begin{table}[htbp]
	\caption{Words used in crowd accidents reports ranked based on the binary count approach and by considering occurrences in each geographical area. This table is equivalent to the representation of \figurename~\ref{fig:binary_area_occurrence}.}
	\label{tab:area_binary_occurrences}
	\begin{tabular}{lllll}
		\hline
			 & Europe         & India \& Pakistan & USA           & Other         \\
		\hline
		1  & person (134)   & stampede (93)     & person (76)   & stampede (49) \\
		2  & stampede (123) & person (70)       & stampede (71) & person (45)   \\
		3  & hospital (73)  & incident (57)     & crowd (46)    & hospital (35) \\
		4  & police (70)    & hospital (51)     & death (41)    & death (29)    \\
		5  & crowd (67)     & woman (50)        & official (40) & incident (24) \\
		6  & incident (66)  & place (48)        & victim (38)   & city (23)     \\
		7  & official (65)  & police (46)       & hospital (35) & crowd (22)    \\
		8  & death (63)     & thousand (36)     & stadium (33)  & official (22) \\
		9  & thousand (57)  & crowd (35)        & police (32)   & crush (17)    \\
		10 & crush (54)     & lakh (30)         & child (30)    & hundred (17)  \\
		\hline
	\end{tabular}
\end{table}
\begin{table}[htbp]
	\caption{Words used in crowd accidents reports ranked based on the binary count approach and by considering total link strength in each geographical area. This table is equivalent to the representation of \figurename~\ref{fig:binary_area_strength}.}
	\label{tab:area_binary_strength}
	\begin{tabular}{lllll}
		\hline
			 & Europe          & India \& Pakistan & USA             & Other          \\
		\hline
		1  & person (5095)   & stampede (2311)   & person (2116)   & stampede (902) \\
		2  & stampede (4099) & person (1887)     & stampede (1973) & person (854)   \\
		3  & police (3183)   & incident (1590)   & crowd (1484)    & hospital (730) \\
		4  & crowd (2955)    & hospital (1450)   & death (1292)    & death (640)    \\
		5  & hospital (2830) & place (1432)      & victim (1261)   & city (475)     \\
		6  & death (2738)    & woman (1292)      & official (1235) & crowd (469)    \\
		7  & incident (2691) & police (1211)     & hospital (1083) & incident (467) \\
		8  & year (2554)     & crowd (1076)      & city (1055)     & official (439) \\
		9  & crush (2267)    & thousand (1074)   & police (1042)   & year (356)     \\
		10 & time (2249)     & lakh (1028)       & year (1018)     & crush (346)    \\
		\hline
	\end{tabular}
\end{table}
\begin{table}[htbp]
	\caption{Words used in crowd accidents reports ranked based on the full count approach and by considering occurrences in each geographical area. This table is equivalent to the representation of \figurename~\ref{fig:full_area_occurrence}.}
	\label{tab:area_full_occurrences}
	\begin{tabular}{lllll}
		\hline
			 & Europe         & India \& Pakistan & USA            & Other          \\
		\hline
		1  & person (639)   & stampede (395)    & person (315)   & person (197)   \\
		2  & stampede (390) & person (254)      & stampede (208) & stampede (163) \\
		3  & police (174)   & incident (128)    & fan (103)      & student (80)   \\
		4  & stadium (159)  & woman (120)       & crowd (99)     & hospital (59)  \\
		5  & fan (152)      & hospital (102)    & stadium (95)   & official (52)  \\
		6  & crowd (130)    & police (88)       & official (79)  & stadium (51)   \\
		7  & year (118)     & devotee (87)      & child (75)     & crowd (50)     \\
		8  & hospital (112) & pilgrim (86)      & police (73)    & school (50)    \\
		9  & death (104)    & temple (86)       & death (68)     & death (46)     \\
		10 & incident (104) & place (74)        & victim (64)    & incident (39)  \\
		\hline
	\end{tabular}
\end{table}
\begin{table}[htbp]
	\caption{Words used in crowd accidents reports ranked based on the full count approach and by considering total link strength in each geographical area. This table is equivalent to the representation of \figurename~\ref{fig:full_area_strength}.}
	\label{tab:area_full_strength}
	\begin{tabular}{lllll}
		\hline
			 & Europe              & India \& Pakistan & USA              & Other           \\
		\hline
		1  & person (81016)      & stampede (22443)  & person (26432)   & person (11753)  \\
		2  & duckenfield (37433) & person (19542)    & stampede (13873) & stampede (6500) \\
		3  & officer (34847)     & pilgrim (11023)   & crowd (10142)    & crowd (5099)    \\
		4  & police (33927)      & incident (7906)   & fan (9336)       & student (4499)  \\
		5  & fan (30619)         & crowd (7203)      & door (8285)      & jess (4160)     \\
		6  & year (29588)        & hospital (7075)   & official (8242)  & band (3660)     \\
		7  & time (22857)        & woman (7060)      & game (7544)      & liza (3587)     \\
		8  & stampede (22812)    & police (6946)     & police (7291)    & hospital (3324) \\
		9  & evidence (22107)    & temple (6520)     & victim (6623)    & school (3101)   \\
		10 & force (21999)       & time (6304)       & death (6520)     & year (2871)     \\
		\hline
	\end{tabular}
\end{table}

\subsection*{Purpose of gathering}
Tables in this section have been rotated compared to the previous representation given the large number of categories and the reduced extent of the ranking.
\begin{table}[htbp]
	\footnotesize
	\caption{Words used in crowd accidents reports ranked based on the binary count approach and by considering occurrences for different purposes of gathering. This table is equivalent to the representation of \figurename~\ref{fig:binary_event_occurrence}.}
	\label{tab:event_binary_occurrences}
	\begin{tabular}{llllll}
		\hline
										& 1              & 2             & 3             & 4             & 5                 \\
		\hline
		Application     & stampede (8)   & gate (7)      & ground (6)    & person (6)    & place (6)         \\
		Donation        & stampede (24)  & person (19)   & woman (17)    & agency (11)   & distribution (11) \\
		Educational     & school (21)    & stampede (21) & hospital (18) & student (18)  & child (15)        \\
		Entertainment   & person (66)    & stampede (61) & hospital (42) & crowd (35)    & incident (35)     \\
		Political       & stampede (22)  & person (21)   & hospital (17) & rally (16)    & election (15)     \\
		Religious       & stampede (113) & person (106)  & woman (66)    & thousand (61) & incident (54)     \\
		Shopping        & person (11)    & stampede (10) & crowd (9)     & sale (8)      & shopper (8)       \\
		Sport           & person (67)    & stadium (65)  & fan (57)      & match (57)    & game (48)         \\
		Transportation  & stampede (22)  & station (20)  & train (18)    & person (17)   & passenger (14)    \\
		Other / Unknown & person (8)     & stampede (8)  & building (5)  & police (5)    & polouse (5)       \\
		\hline
	\end{tabular}
\end{table}
\begin{table}[htbp]
	\footnotesize
	\caption{Words used in crowd accidents reports ranked based on the binary count approach and by considering total link strength for different purposes of gathering. This table is equivalent to the representation of \figurename~\ref{fig:binary_event_strength}.}
	\label{tab:event_binary_strength}
	\begin{tabular}{llllll}
		\hline
										& 1               & 2               & 3               & 4                      & 5                  \\
		\hline
		Application     & stampede (46)   & gate (43)       & ground (38)     & recruitment drive (38) & person (35)        \\
		Donation        & stampede (245)  & person (200)    & woman (200)     & agency (135)           & distribution (129) \\
		Educational     & school (355)    & stampede (355)  & student (318)   & hospital (302)         & child (263)        \\
		Entertainment   & person (1714)   & stampede (1522) & hospital (1185) & crowd (1084)           & incident (928)     \\
		Political       & stampede (411)  & person (403)    & hospital (329)  & election (324)         & rally (323)        \\
		Religious       & stampede (3617) & person (3480)   & thousand (2201) & woman (2124)           & crowd (1975)       \\
		Shopping        & person (75)     & stampede (71)   & crowd (65)      & shopper (58)           & door (53)          \\
		Sport           & person (2479)   & stadium (2325)  & fan (2203)      & match (2090)           & police (1919)      \\
		Transportation  & stampede (283)  & station (273)   & train (254)     & person (247)           & passenger (191)    \\
		Other / Unknown & person (33)     & stampede (33)   & police (25)     & polouse (25)           & witness (25)       \\
		\hline
	\end{tabular}
\end{table}
\begin{table}[htbp]
	\footnotesize
	\caption{Words used in crowd accidents reports ranked based on the full count approach and by considering occurrences for different purposes of gathering. This table is equivalent to the representation of \figurename~\ref{fig:full_event_occurrence}.}
	\label{tab:event_full_occurrences}
	\begin{tabular}{llllll}
		\hline
										& 1               & 2              & 3             & 4             & 5                      \\
		\hline
		Application     & university (21) & ground (20)    & stampede (20) & person (16)   & recruitment drive (16) \\
		Donation        & person (91)     & stampede (79)  & woman (38)    & camp (22)     & agency (21)            \\
		Educational     & student (124)   & school (76)    & stampede (71) & child (40)    & accident (33)          \\
		Entertainment   & person (350)    & stampede (161) & concert (80)  & crowd (79)    & hospital (72)          \\
		Political       & person (104)    & stampede (89)  & rally (53)    & incident (34) & woman (34)             \\
		Religious       & stampede (461)  & person (421)   & pilgrim (184) & woman (156)   & temple (141)           \\
		Shopping        & person (39)     & store (30)     & stampede (28) & ticket (23)   & sale (21)              \\
		Sport           & person (286)    & fan (234)      & stadium (219) & match (152)   & stampede (143)         \\
		Transportation  & stampede (93)   & person (68)    & station (49)  & train (41)    & official (37)          \\
		Other / Unknown & person (41)     & stampede (20)  & panic (18)    & station (16)  & police (14)            \\
		\hline
	\end{tabular}
\end{table}
\begin{table}[htbp]
	\footnotesize
	\caption{Words used in crowd accidents reports ranked based on the full count approach and by considering total link strength for different purposes of gathering. This table is equivalent to the representation of \figurename~\ref{fig:full_event_strength}.}
	\label{tab:event_full_strength}
	\begin{tabular}{llllll}
		\hline
										& 1                & 2                & 3                   & 4               & 5                   \\
		\hline
		Application     & university (861) & gate (608)       & stampede (544)      & ground (542)    & candidate (538)     \\
		Donation        & person (2784)    & stampede (2385)  & woman (1319)        & camp (1223)     & refugee (1086)      \\
		Educational     & student (5302)   & school (3346)    & stampede (2893)     & accident (1579) & child (1423)        \\
		Entertainment   & person (26980)   & crowd (9066)     & stampede (8157)     & hospital (5401) & jess (5060)         \\
		Political       & person (5041)    & stampede (4235)  & rally (2654)        & incident (1934) & jonathan (1897)     \\
		Religious       & person (33276)   & stampede (29426) & pilgrim (19389)     & woman (10335)   & crowd (10320)       \\
		Shopping        & store (1338)     & crowd (935)      & person (912)        & worker (795)    & shopper (765)       \\
		Sport           & person (56648)   & fan (35729)      & duckenfield (34973) & officer (32837) & police (29439)      \\
		Transportation  & stampede (4060)  & person (3753)    & station (2495)      & official (2016) & train (1956)        \\
		Other / Unknown & person (2291)    & panic (1209)     & station (1166)      & polouse (925)   & oxford street (857) \\     
		\hline
	\end{tabular}
\end{table}

\section*{Appendix C}
This appendix contains numerical values for the results from the sentiment analysis. The presentation follows the same order of the main text. Love and surprise have been excluded to save space considering their very marginal relevance (only in a very few cases their score is higher than 0.01 and typically 99\% of the total score is given by fear, anger, sadness, and joy).

\subsection*{Historical trend}
\begin{table}[htbp]
	\caption{Sentiment score associated with crowd accidents reported during different time periods depending on the analyzed part of the report. This table is equivalent to the representation of \figurename~\ref{fig:sentiment_time}.}
	\label{tab:sentiment_time_all}
	\begin{tabular}{lllllllll}
		\hline
		Time period		& \multicolumn{2}{c}{Fear}	& \multicolumn{2}{c}{Anger} & \multicolumn{2}{c}{Sadness}	& \multicolumn{2}{c}{Joy} 	\\
									& Body 		& Title 					& Body 		& Title						& Body 		& Title 						& Body 		& Title 					\\
		\hline
		1900--1979 		& 0.554 	& 0.500 					& 0.139 	& 0.333 					& 0.258   & 0.167   					& 0.023 	& 0.000 					\\
		1980--1999 		& 0.549 	& 0.654 					& 0.172 	& 0.154 					& 0.122   & 0.154   					& 0.117 	& 0.038 					\\
		2000--2009 		& 0.669 	& 0.650 					& 0.126 	& 0.225 					& 0.097   & 0.100   					& 0.090 	& 0.025 					\\
		2010--2014 		& 0.584 	& 0.733 					& 0.137 	& 0.128 					& 0.134   & 0.093   					& 0.110 	& 0.047 					\\
		2015--2019 		& 0.616 	& 0.791 					& 0.154 	& 0.087 					& 0.124   & 0.113   					& 0.082 	& 0.009 					\\
		\hline
	\end{tabular}
\end{table}

\subsection*{Geographical origin}
\begin{table}[htbp]
	\caption{Sentiment score associated with crowd accidents reported from institutions in different geographical areas depending on the analyzed part of the report. This table is equivalent to the representation of \figurename~\ref{fig:sentiment_area}.}
	\label{tab:sentiment_area_all}
	\begin{tabular}{lllllllll}
		\hline
		Geographical				& \multicolumn{2}{c}{Fear}	& \multicolumn{2}{c}{Anger} & \multicolumn{2}{c}{Sadness}	& \multicolumn{2}{c}{Joy} 	\\
		area								& Body 		& Title 					& Body 		& Title						& Body 		& Title 					& Body 		& Title 						\\
		\hline
		Europe             	& 0.588 	& 0.588 					& 0.140 	& 0.228 					& 0.142 	& 0.154 					& 0.111 	& 0.029 						\\
		India \& Pakistan 	& 0.638 	& 0.900 					& 0.129 	& 0.022 					& 0.107 	& 0.056 					& 0.094 	& 0.022 						\\
		USA                	& 0.621 	& 0.688 					& 0.167 	& 0.182 					& 0.104 	& 0.104 					& 0.074 	& 0.026 						\\
		Other              	& 0.676 	& 0.760 					& 0.130 	& 0.140 					& 0.104 	& 0.080 					& 0.073 	& 0.020							\\
		\hline
	\end{tabular}
\end{table}

\subsection*{Purpose of gathering}
Tables in this section have been rotated compared to the previous representation given the large number of categories and the reduced number of sentiments.
\begin{table}[htbp]
	\caption{Sentiment score associated with the reporting of crowd accidents occurring in different circumstances (or gathering purposes) and depending on the analyzed part of the report. This table complement with numerical data the representation of \figurename~\ref{fig:sentiment_event}.}
	\label{tab:sentiment_event_all}
	\begin{tabular}{lllllllll}
		\hline
		Purpose of			& \multicolumn{2}{c}{Fear}	& \multicolumn{2}{c}{Anger} & \multicolumn{2}{c}{Sadness}	& \multicolumn{2}{c}{Joy} 	\\
		gathering				& Body 		& Title 					& Body 		& Title						& Body 		& Title 					& Body 		& Title 						\\
		\hline
		Application    	& 0.576 	& 0.875 					& 0.217 	& 0.000 					& 0.083 	& 0.125 					& 0.088 	& 0.000 						\\
		Donation       	& 0.600 	& 0.708 					& 0.153 	& 0.125 					& 0.119 	& 0.083 					& 0.107 	& 0.083							\\
		Educational    	& 0.788 	& 0.850 					& 0.104 	& 0.150 					& 0.057 	& 0.000 					& 0.040 	& 0.000							\\
		Entertainment  	& 0.609 	& 0.492 					& 0.148 	& 0.262 					& 0.107 	& 0.215 					& 0.106 	& 0.031							\\
		Political      	& 0.572 	& 0.905 					& 0.173 	& 0.095 					& 0.138 	& 0.000 					& 0.076 	& 0.000							\\
		Religious      	& 0.616 	& 0.770 					& 0.109 	& 0.097 					& 0.133 	& 0.106 					& 0.121 	& 0.027							\\
		Shopping       	& 0.555 	& 0.667 					& 0.134 	& 0.167 					& 0.140 	& 0.083 					& 0.149 	& 0.083							\\
		Sport          	& 0.577 	& 0.617 					& 0.204 	& 0.233 					& 0.142 	& 0.133 					& 0.059 	& 0.017							\\
		Transportation 	& 0.703 	& 1.000 					& 0.088 	& 0.000 					& 0.094 	& 0.000 					& 0.048 	& 0.000							\\
		Other/Unknown  	& 0.773 	& 0.750 					& 0.139 	& 0.250 					& 0.044 	& 0.000 					& 0.035 	& 0.000							\\
		\hline
	\end{tabular}
\end{table}

\section*{Appendix D}
Position of ``panic'' within the ranking presented in \figurename~\ref{fig:words_time}, \figurename~\ref{fig:words_area}, and \figurename~\ref{fig:words_event}. For each attribute (time period, geographical area, and purpose of gathering) the absolute position within the ranking and relative position are given. The relative position represents the position of ``panic'' divided by the total number of entries. For each graph the vertical axis is inverted to have higher positions to appear at the top.

\begin{figure}[htbp]
		\centering
		\subfigure[{Absolute position} 	\label{fig:panic_time_1}]
		{\includegraphics[width=60mm]{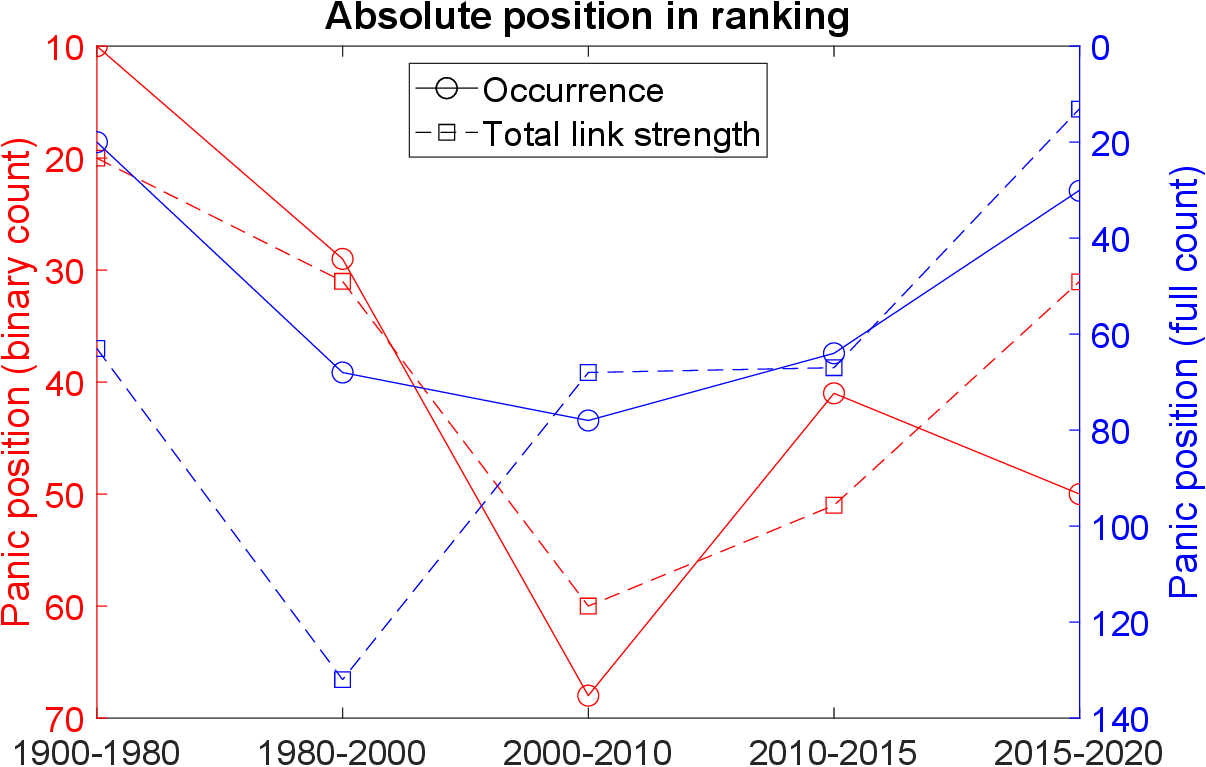}}
		\subfigure[{Relative position} 	\label{fig:panic_time_2}]
		{\includegraphics[width=60mm]{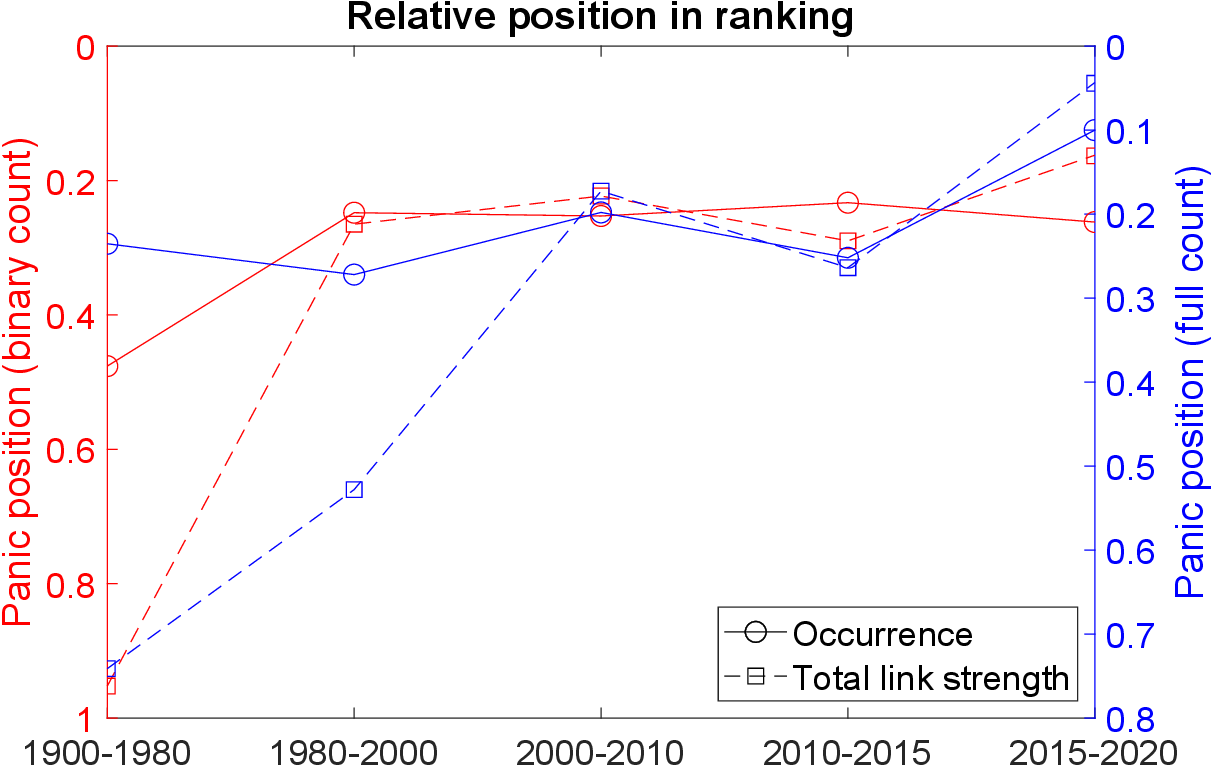}}
    \caption{Change in the position of ``panic'' within text corpus representing different time periods. As the graphs show, ``panic'' is consistently in the ranking, and when trends are examined, it appears to be partially regaining popularity (this trend is clearer when viewed in relative terms).}
		\label{fig:panic_time}
\end{figure}

\begin{figure}[htbp]
		\centering
		\subfigure[{Absolute position} 	\label{fig:panic_area_1}]
		{\includegraphics[width=60mm]{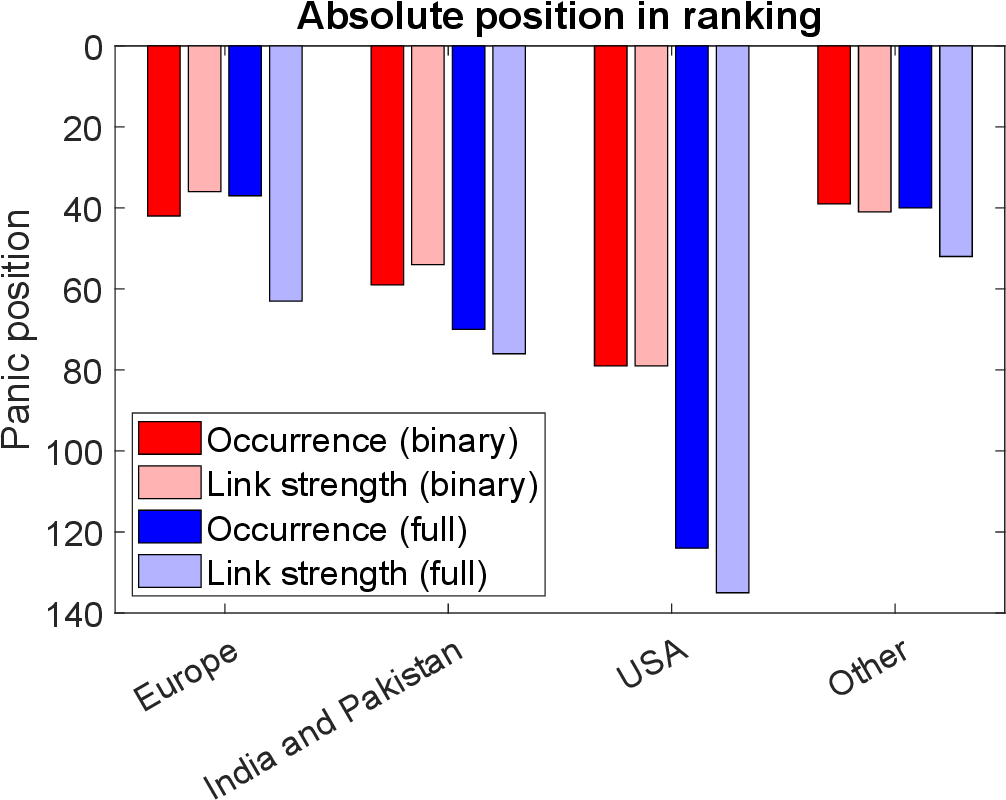}}
		\subfigure[{Relative position} 	\label{fig:panic_area_2}]
		{\includegraphics[width=60mm]{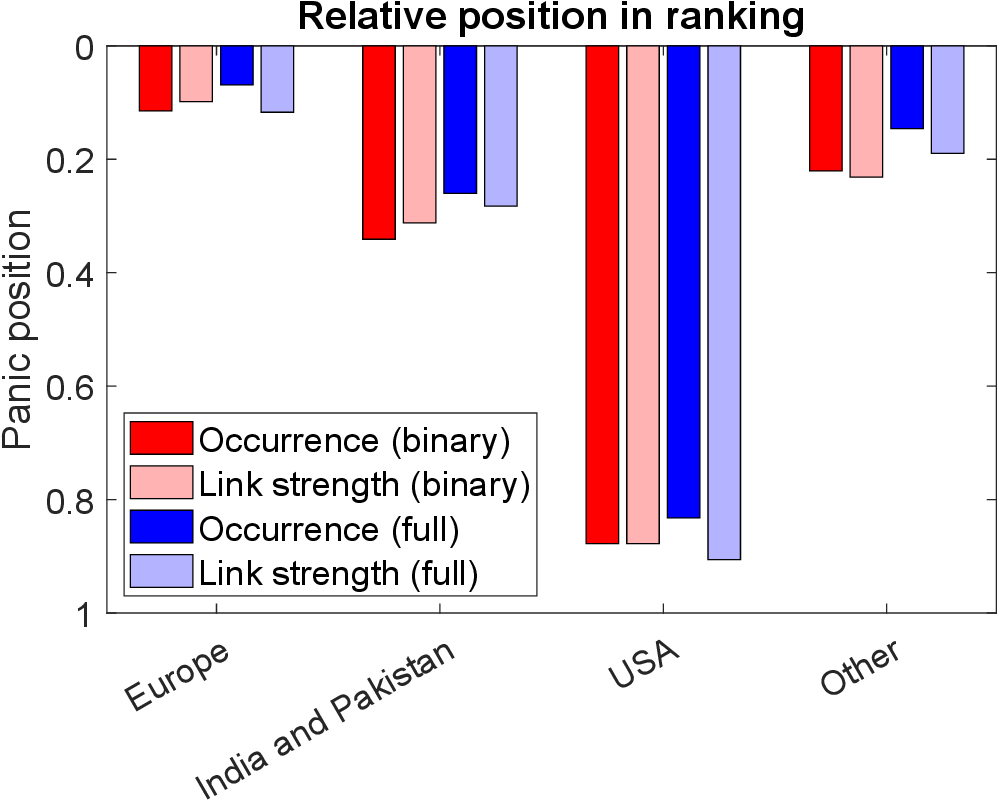}}
    \caption{Use of ``panic'' in different geographical areas (based on the reporting source). Europe and other areas appear to use ``panic'' more frequently compared to the USA in which low positions are often found in the ranking.}
		\label{fig:panic_area}
\end{figure}

\begin{figure}[htbp]
		\centering
		\subfigure[{Absolute position} 	\label{fig:panic_event_1}]
		{\includegraphics[width=60mm]{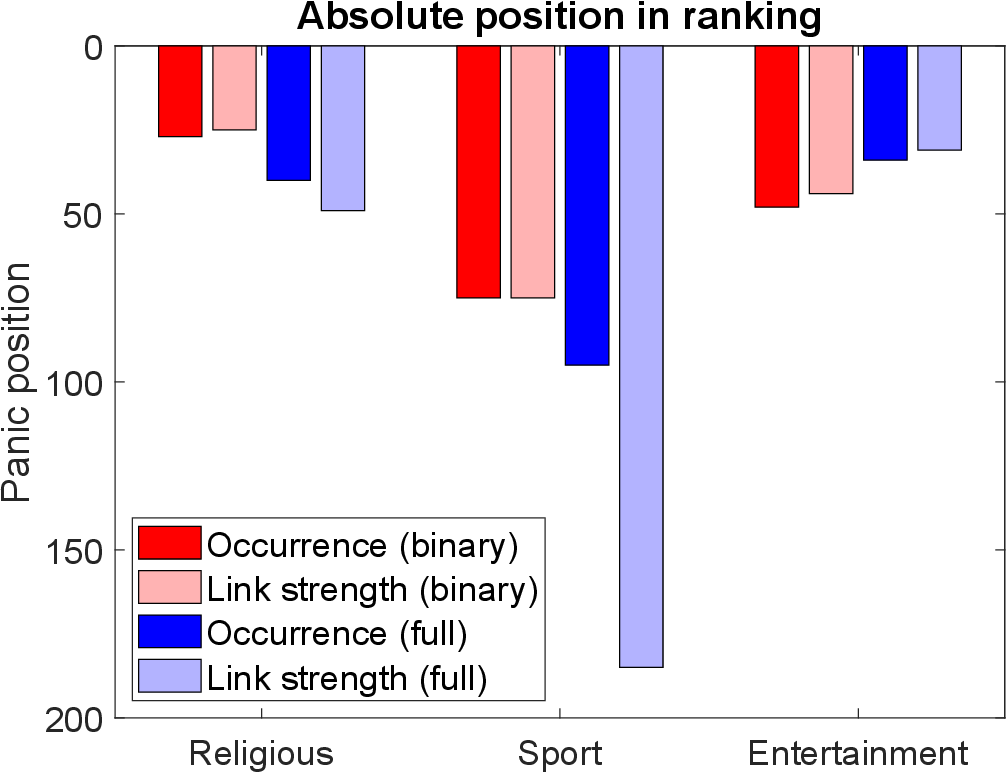}}
		\subfigure[{Relative position} 	\label{fig:panic_event_2}]
		{\includegraphics[width=60mm]{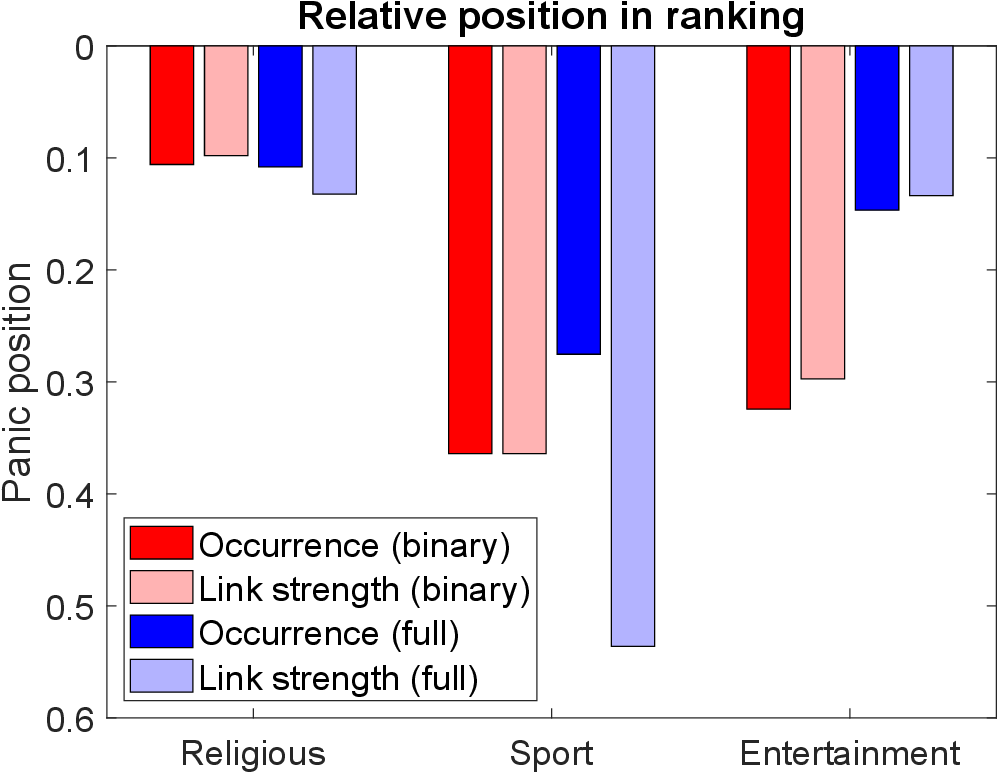}}
    \caption{Use of ``panic'' when reporting crowd accidents occurring in different gathering contexts. Like for \figurename~\ref{fig:sentiment_event}, the purposes of gathering having too little sample size have been excluded focusing on religious, sport, and entertainment events. ``Panic'' appears to be used more frequently for religious events and, to a lesser extent, in sport and entertainment events.}
		\label{fig:panic_event}
\end{figure}

\bibliographystyle{abbrvnat}
\bibliography{crowd_accident_reporting}

\end{document}